\documentclass[a4paper, 11pt]{article}

\usepackage{preamble}

\title{Invariant relativistic kinematics}
\subtitle{Phase space triangulation}

\author{Jan Hajer \email{jan.hajer@tecnico.ulisboa.pt}}

\AtBeginDocument{\hypersetup{pdfauthor=Jan Hajer}}
\affiliation{Centro de Física Teórica de Partículas (CFTP), Instituto Superior Técnico (IST), Universidade de Lisboa, Av.\ Rovisco Pais 1, 1049-001 Lisboa, Portugal}

\begin{document}

\maketitle

\begin{abstract}
The calculation of particle decay widths and scattering cross sections naturally decomposes into a quantum mechanical amplitude and a relativistic \PS.
This \PS can be formulated in terms of parallelotopes providing frame independent invariants.
We demonstrate how these invariants are related to frame dependent observables such as momenta, energies, and angles between particles.
Furthermore, we derive expressions for $n$-dimensional \PSs featuring simple integration limits that are particularly well suited for an analytical treatment.
To that end we develop a pictorial description using \PS diagrams that allow to straightforwardly identify the optimal set of integration variables for arbitrary $n$.
\end{abstract}

\tableofcontents
\listoffigures
\clearpage

\section{Introduction}

The calculation of scattering cross sections \cite{Rutherford:1911,Mott:1929} and decay widths \cite{Weisskopf:1930au} is a central application of quantum field theory in particle physics \cite{Fermi:1932xva}.
These calculations naturally separate into two parts: the derivation of the quantum mechanical amplitude and the relativistic \PS.
Because the amplitude encodes the essence of the particle interaction, most publications emphasise its calculation over that of the \PS.
In addition, an intuitive pictorial representation in terms of Feynman diagrams is well-established \cite{Feynman:1948km, Feynman:1949zx, Feynman:1949hz}.

After an early period of intensive research on the parameterisation of differential \PSs \cite{Byers:1964ryc, Byckling:1969sx, Kumar:1969jjy, Kumar:1970wm, Morrow:1970bh, Poon:1970zv, Byckling:1973yx}, interest in further development of the topic declined and the kinematics of particle interactions and the construction of \PSs are now typically treated in standard textbooks \cite{Hagedorn:1963hdh, Kallen:1964lxa, Byckling:1971vca, Kinematics:2022pth}.
Although there have been attempts to develop diagrammatic tools for \PS calculations \cite{Asribekov:1962tgp,Jing:2020tth}, none of these approaches has become widely adopted.

In this work, we aim to bridge this gap by developing a formulation of many-body \PSs using only invariant quantities.
Naturally, this calculation recovers known results \cite{Byckling:1971vca}.
Beyond these expressions we introduce a parameterisation of the \PS in terms of angular variables with simple integration limits that is particularly well suited for analytic calculations.
To this end, we introduce a pictorial description based on \PS diagrams, which encode the essential kinematic information of particle interactions in a geometric and frame-independent manner.

We begin in \cref{sec:PS} by introducing the relativistic \PS and provide expressions for the particle decay width and the scattering cross section.
We follow up in \cref{sec:distance geometry} by employing determinants used in distance geometry to compute volumes and angles of parallelotopes in Euclidean space as well as \MST and relate them to invariants in particle interactions.
After establishing these basics we show in \cref{sec:two-body interaction} how the invariant area of a parallelogram can be used to compute both the frame-dependent energies and momenta of particle interactions, as well as the two-body \PS.
Subsequently, we relate the volume of a parallelepiped to the opening angle between interacting particles and compute the three-body \PS in \cref{sec:three-body interaction}.
Afterwards we connect the volume of a four-dimensional parallelotope to the angle between two decay planes and compute the four-body \PS in \cref{sec:four-body interaction}.
In \cref{sec:five-body interaction}, we show how the volume of a five-dimensional parallelotope leads to a constraint relevant for the calculation of the five-body \PS.
This derivation culminates in the generalisation of this construction to arbitrary-dimensional \PSs in \cref{sec:multi-body interaction}.
Finally, we conclude in \cref{sec:conculsion}.

\section{\texorpdfstring{\sentence\PSlong}{Phase space}} \label{sec:PS}

\resetacronym{PS}

The differential \PS of a relativistic $n$-body interaction is defined as a function of the four-momenta $\vec4 p_i = (E_i, \vec3 p_i)$ via
\footnote{
We indicate suppressed three- and four-space indices using $\vec3 p$ and $\vec4 p$, respectively.
}
\begin{equation} \label{eq:momentum PS}
\d^{3n-4} \Phi_n(\vec4 p_{a\cdots n}; \vec4 p_a, \dots, \vec4 p_n) = \zeta \vec4 p_{a\cdots n}^2 \iiiint \diractwopi^4\left(\vec4 p_{a\cdots n} - \textstyle\sum\nolimits_{i=a}^n \vec4 p_i\right) \prod_{i=a}^n \int \diractwopi(\zeta \vec4 p_i^2 - m_i^2) \dtwopi^4 \vec4 p_i ,
\end{equation}
where the Dirac distributions place the outgoing particles on-shell and ensure energy-momentum conservation by forcing one of the momenta to be equal to the sum $\vec4 p_{a\cdots n}^{} = \vec4 p_a + \dots + \vec4 p_n$.
The sign of the momentum squares depends on the signature of the metric
\footnote{
The pseudo-Euclidean metric contains an overall minus sign in comparison to the Euclidean metric and appears \eg after a Wick rotation of both of the Lorentzian metrics.
}
\begin{equation} \label{eq:metric signature}
\zeta = \begin{cases}
1 & \text{Euclidean space and \MST with mostly negative metric,} \\
-1 & \text{pseudo-Euclidean space and \MST with mostly positive metric,}
\end{cases}
\end{equation}
and the barred quantities are normalised with powers of $2\pi$
\begin{align} \label{eq:barred normalisation}
\diractwopi^n (x-y) &= (2\pi)^n \dirac^n (x-y) , &
\dtwopi^n x &= \frac{\d^n x}{(2\pi)^n}.
\end{align}
Since the differential \PS can be expressed in terms of invariant masses \cite{Byckling:1969sx,Morrow:1970bh,Poon:1970zv} we write
\begin{equation} \label{eq:mass momentum PS}
\d^{3n-4} \Phi_n(m_{a\cdots n}; m_a, \dots, m_n) := \d^{3n-4} \Phi_n(\vec4 p_{a\cdots n}; \vec4 p_a, \dots, \vec4 p_n).
\end{equation}
In order to facilitate a generic notation we use only invariant mass squares
\begin{align} \label{eq:invariant mass}
\m_i^2 &:= \zeta \vec4 p_i^2,
\end{align}
in place of Mandelstam variables \cite{Mandelstam:1958xc}.
In the differential \PS \eqref{eq:mass momentum PS} they are fixed to the on-shell masses $m_i$ by the Dirac distributions.
In the following we abbreviate the arguments of functions~$f$ that depend on $n-i$ independent particles using
\begin{equation} \label{eq:multi-body notation}
f(m_{a\cdots i\cdots n}) := f(\m_{a\cdots i}, m_{i+1}, \dots, m_n) := f(m_{a\cdots i\cdots n}; \m_{a\cdots i}, m_{i+1}, \dots, m_n) ,
\end{equation}
when no disambiguity can arise.
Furthermore, the relative orientation of the differential \PS \eqref{eq:mass momentum PS} with respect to a reference frame is relevant.
In \cref{sec:solid angle differential} we show that the relative orientation of the two-body \PS can be defined by the solid angle differential \eqref{eq:solid angle differential} by adding two additional particles with masses $m_{n+1}$ and $m_{n+2}$.
This argument is summarised in \cref{fig:solid angle definition}.
As shown in \cref{sec:three-body PS} the relative orientation of higher dimensional \PSs is defined by the Euler angle differential \eqref{eq:Euler angle differential} that depends on a third additional particle with mass $m_{n+3}$.
Therefore, we write
\begin{equation}
\d^{3n-4} \Phi_n(m_{a\cdots i\cdots n}; m_{n+1}, \dots) := \d^{3n-4} \Phi_n(m_{a\cdots i\cdots n}; m_{a\cdots i}, m_{i+1}, \dots, m_n; m_{n+1}, \dots) .
\end{equation}
A high-dimensional differential \PS can be generated from two lower-dimensional differential \PSs using the recursion relation \cite{Srivastava:1958ve}
\footnote{
Where we have used that $\delta(a-b) = \int \delta(a-x) \delta(x-b) \d x$.
}
\begin{multline} \label{eq:recursion relation}
\d^{3n-4} \Phi_n(m_{a\cdots n}; m_{n+1}, \dots) = 2 \frac{\dfourpi \m_{a\cdots i}^2}{\m_{a\cdots i}^2} \d^{3i-4} \Phi_i(\m_{a\cdots i}; m_{i+1}, \dots) \\
\d^{3(n-i)-1} \Phi_{n-i+1}(m_{a\cdots n}; \m_{a\cdots i}^{}, m_{i+1}, \dots, m_n; m_{n+1}, \dots) ,
\end{multline}
where the tilded quantities are normalised with powers of $4\pi$
\begin{align} \label{eq:tilded normalisation}
\diracfourpi^n (x-y) &= (4\pi)^n \dirac^n (x-y) , &
\dfourpi^n x &= \frac{\d^n x}{(4\pi)^n} .
\end{align}
In particular we rely on the recursive factorisation into differential two-body \PSs via
\begin{multline} \label{eq:recursion relation two-body}
\d^{3n-4} \Phi_n(m_{a\cdots n}; m_{n+1}, \dots ) = \\
\d^3 \Phi_2^\prime(\m_{ab}; m_c, m_d) \d^{3n-7} \Phi_{n-1}(m_{a\cdots n}; \m_{ab}^{}, m_c, \dots, m_n; m_{n+1}, \dots) ,
\end{multline}
where we have defined the augmented two-body \PS differential
\begin{equation} \label{eq:augmented two-body PS differential}
\d^3 \Phi_2^\prime(\m_{ab}; m_c, m_d) = 2 \frac{\dfourpi \m_{ab}^2}{\m_{ab}^2} \d^2 \Phi_2(\m_{ab}; m_c, m_d) .
\end{equation}
The explicit frame independent expression for the differential two-body \PS is derived in \cref{sec:two-body PS} and given in \eqref{eq:differential two-body PS}.
Two different parameterisations of the augmented two-body \PS differential are depicted in \cref{fig:augmented}.

\subsection{Decay width and cross section}

We define the differential $n$-body quantum \PS as the product of the squared amplitude and the differential $n$-body \PS
\begin{equation} \label{eq:quantum PS}
\d^{3n-4} \Psi_n(m_{a\cdots n}; m_{n+1}, \dots) = \frac12 \abs{\mathcal A_{n+1}}^2 \d^{3n-4} \Phi_n(m_{a\cdots n}; m_{n+1}, \dots) .
\end{equation}
Note that four-momentum conservation restricts the momenta of an $n+1$ particle interaction to an $n$-body \PS.
Since the mass-dimensions of the amplitude and the \PS are $3-n$ and $2n-2$, respectively, the quantum \PS has mass-dimension four.
The differential decay width of a particle with mass $m_{a\cdots n}$ decaying into $n$ particles is then given by \cite{Kinematics:2022pth}
\begin{equation} \label{eq:decay width}
\d^{3n-4} \Gamma_n(m_{a\cdots n}; m_{n+1}, \dots) = \frac{\d^{3n-4} \Psi_n(m_{a\cdots n}; m_{n+1}, \dots)}{\energy_{a\cdots n}(m_{a\cdots n+1}) m_{a\cdots n}^2} ,
\end{equation}
and has mass dimension one.
The energy $\energy_{a\cdots n}(m_{a\cdots n+1})$ is measured in the rest frame of the observer, defined here as the rest frame of the particle with mass $m_{a\cdots n+1}$.
In the rest frame of the decaying particle $m_{a\cdots n+1} \equiv m_{a\cdots n}$ the energy becomes $\energy_{a\cdots n}(m_{a\cdots n}) = m_{a\cdots n}$.
See also the discussion around the identities \eqref{eq:energy momentum special cases}.

We define the differential cross section of two particles with masses $m_{a\cdots n}$ and $m_n$ scattering into $n-1$ particles by
\begin{multline} \label{eq:cross section}
\d^{3n-5} \sigma_n(m_{a\cdots n}, m_n; m_{a\cdots n-1}; m_a, \dots, m_{n-1}; m_{n+1}, \dots) = \\
\pi \int \diracfourpi(\m_{a\cdots n-1}^2 - m_{a\cdots n-1}^2) \frac{\d^{3n-4} \Psi_n(m_{a\cdots n}; m_{n+1}, \dots)}{\volume_2^2(m_{a\cdots n})} ,
\end{multline}
where the Dirac distribution fixes the invariant mass $\m_{a\cdots n-1}$ to the \COM energy $m_{a\cdots n-1}$.
The square of the area $\volume_2$ will be introduced in \cref{sec:triangles} and is related to the commonly used Källén triangle function \cite{Kallen:1964lxa} via $\volume_2^2(m_{a\cdots n}) = \flatfrac{\lambda(m_{a\cdots n}^2, \m_{a\cdots n-1}^2, m_n^2)}{4}$.
Since it has mass dimension four the mass dimension of the cross section is minus two.
\footnote{
For practical purpose it is useful to know that $\flatfrac{\hbar^2 c^2}{\unit{GeV^2}} \approx \unit[0.3894]{mb}$.
}
Note that the momentum of the particle with mass $m_n$ is reversed in comparison to the notation used in the definition of the \PS \eqref{eq:momentum PS} and the decay width \eqref{eq:decay width}.
This fact becomes important when calculating quantities that depend on the Gram determinant \eqref{eq:Gram determinant}, such as the extremal values \eqref{eq:three-body integration limits ingredients} or the cosine of the polar opening angle \eqref{eq:three-body cosine} and the cosine of the azimuthal decay plane angle \eqref{eq:four-body cosine}.
Using the recursion relation \eqref{eq:recursion relation two-body} the $n$-body \PS of the cross section can be split into an augmented two-body \PS differential \eqref{eq:augmented two-body PS differential} for the incoming particles and an outgoing $n-1$-body \PS.
After inserting the explicit expression for the two-body \PS \eqref{eq:integrated two-body PS} of the incoming particles and integrating over the corresponding solid angle differential \eqref{eq:solid angle differential integrated} as well as the invariant mass differential that corresponds to the \COM energy the differential cross sections reads
\begin{multline} \label{eq:cross section integrated}
\d^{3n-7} \sigma_n(m_{a\cdots n}, m_n; m_{a\cdots n-1}; m_a, \dots, m_{n-1}; m_{n+1}, \dots) = \\
\abs{\mathcal A_{n+1}}^2 \frac{\d^{3n-7} \Phi_{n-1}(m_{a\cdots n-1}; m_n, \dots)}{4 \volume_2(m_{a\cdots n}) m_{a\cdots n-1}^2} ,
\end{multline}
where the amplitude contains the complete process, while the differential \PS describes only the outgoing part of the process.
Furthermore, only one of the integrals over the external Euler angle differential \eqref{eq:Euler angle differential} remains in this expressions and the seemingly external angle differentials in the \PS of the final state particles is connected to the incoming particles.
The factor $4 \volume_2(m_{a\cdots n})$ in the cross section \eqref{eq:cross section integrated} is commonly called Møller flux factor \cite{Moller:1945gen,Cannoni:2016hro}.

\section{Distance geometry} \label{sec:distance geometry}

\resetacronym{MST}

Four-momentum conservation constrains the physically accessible \MST in an $n$-body interaction to $n$-dimensional parallelotopes $\parallel_n(m_a, \dots, m_n)$ whose edges are given by the particle masses.
Just as in Euclidean space these parallelotopes can be triangulated by simplices $\simplex_n(m_a, \dots, m_n)$ of the same dimension.
Relevant properties of the particle interaction can be identified with angles of these simplices.
Since the parallelotopes and simplices are completely defined by their edges their properties can be derived using determinants of the four-momenta used in distance geometry.

\subsection{Gram and \CMlong determinants}

Given $2n$ particles with momenta $\vec4 p_n$ we use the $4 \times n$-dimensional matrices $(\vec4 p_a, \vec4 p_b, \dots, \vec4 p_n)$ to define the determinant of the $n$-dimensional Gram matrix \cite{gram1879om}
\begin{equation} \label{eq:Gram determinant vector}
\gram^2_n\begin{pmatrix}\vec4 p_a, \dots, \vec4 p_n \\
\vec4 p_a^\prime, \dots, \vec4 p_n^\prime \end{pmatrix} := \eta^{n+1}
\begin{vmatrix}
\pair{p_a^{}}{p_a^\prime} & \pair{p_a^{}}{p_b^\prime} & \dots & \pair{p_a^{}}{p_n^\prime} \\
\pair{p_b^{}}{p_a^\prime} & \pair{p_b^{}}{p_b^\prime} & \dots & \pair{p_b^{}}{p_n^\prime} \\
\vdots & \vdots & \ddots & \vdots \\
\pair{p_n^{}}{p_a^\prime} & \pair{p_n^{}}{p_b^\prime} & \dots & \pair{p_n^{}}{p_n^\prime} \\
\end{vmatrix} ,
\end{equation}
where the global sign depends on the determinant of the metric
\begin{equation} \label{eq:metric determinant}
\eta = \begin{cases}
1 & \text{Euclidean space}, \\
-1 & \text{\MSTlong} ,
\end{cases}
\end{equation}
and ensures positivity.
The mass dimension of the Gram determinant is $2n$.
In the following we single out two particles with momenta $\vec4 p_a$ and $\vec4 p_b$ and set for all other momenta $\vec4 p^\prime_i = \vec4 p_i^{}$.
Since the result can be expressed in terms of mass squares we introduce the notation
\begin{equation} \label{eq:Gram determinant}
\gram^2_{n-1}(m_a, m_b; m_c, \dots, m_n) := \zeta^{n-1} \gram^2_{n-1}\begin{pmatrix}\vec4 p_a, \vec4 p_c, \dots, \vec4 p_n \\
\vec4 p_b, \vec4 p_c, \dots, \vec4 p_n\end{pmatrix} ,
\end{equation}
where the global sign is given by the signature of the metric \eqref{eq:metric signature}.
Changing the sign of $\vec4 p_a$ or $\vec4 p_b$ as may be necessary for the calculation of the production cross section \eqref{eq:cross section} switches the sign of the Gram determinant.

For $n \geq 2$ the Laplace expansion of the Gram determinant along the first column in terms of minors can be written as
\begin{multline} \label{eq:Laplace expansion Gram determinant}
\eta \gram^2_{n-1}(m_a, m_b; m_c, \dots, m_n) = \\
\gram_1^2(m_a, m_b) \volume^2_{n-2}(m_c, m_d, \dots, m_n) - \sum_{i=c}^n \gram_1^2(m_b, m_i) \gram^2_{n-2}(m_a, m_i; \dots) ,
\end{multline}
where the symmetric Gram determinant
is defined via
\begin{equation} \label{eq:symmetric Gram determinant}
\volume^2_n(m_a, m_b, \dots, m_n) :=
\gram^2_n\begin{pmatrix}m_a, m_a; m_b \dots, m_n\end{pmatrix} .
\end{equation}
It corresponds to the \CM determinant \cite{Cayley:1841, Menger:1928} and can therefore be written as determinant of the $n+2$-dimensional matrix \cite{Regge1964,Eden:1966dnq}
\begin{equation} \label{eq:CM determinant}
\volume^2_n(m_a, \dots, m_n) =
\frac{(-\eta)^{n+1}}{ 2^n}
\begin{vmatrix*}[l]
\header0 & \header1 & \header1 & \header1 & \dots & \header1 \\
\header1 & \header0 & m_a^2 & \m_{ab}^2 & \dots & m_{a\cdots n}^2 \\
\header1 & m_a^2 & \header0 & m_b^2 & \dots & \m_{b\cdots n}^2 \\
\header1 & \m_{ab}^2 & m_b^2 & \header0 & \dots & \m_{c\cdots n}^2 \\
\header \vdots & \header \vdots & \header \vdots &\header \vdots &\ddots &\header \vdots \\
\header1 & m_{a\cdots n}^2 & \m_{b\cdots n}^2 & \m_{c\cdots n}^2 & \dots & \header0
\end{vmatrix*} ,
\end{equation}
where the invariant masses are defined as in \eqref{eq:invariant mass}.
We call invariant masses composed of an uninterrupted and increasing sequence of indices canonical.

\subsection{Invariants in \MSTlong} \label{sec:invariants}

In four-dimensions we define the five contractions of the \LC tensor density with zero to four momenta via
\begin{equation} \label{eq:tensor densities}
\upsilon_{a,\dots,n,\mu_{n+1}\dots\mu_4} = \sqrt{\frac{(\eta \zeta)^n}{n!}} \epsilon_{\mu_1\mu_2\mu_3\mu_4} \vec4 p_a^{\mu_{n+1}} \dots \vec4 p_n^{\mu_4}.
\end{equation}
The products of two different tensor densities with the same number of free indices can then be expressed using the Gram determinants \eqref{eq:Gram determinant vector}
\begin{subequations}
\begin{align}
\upsilon^\prime_{a\nu\rho\sigma} \upsilon^{a\nu\rho\sigma} &= \gram_1^2 \begin{pmatrix}\vec4 p_a\\
\vec4 p_a^\prime \end{pmatrix}, &
\upsilon^\prime_{abc\sigma} \upsilon^{abc\sigma} &= \gram_3^2 \begin{pmatrix}\vec4 p_a, \vec4 p_b, \vec4 p_c \\
\vec4 p_a^\prime, \vec4 p_b^\prime, \vec4 p_c^\prime \end{pmatrix}, \\
\upsilon^\prime_{ab\rho\sigma} \upsilon^{ab\rho\sigma} &= \gram_2^2 \begin{pmatrix}\vec4 p_a, \vec4 p_b \\
\vec4 p_a^\prime, \vec4 p_b^\prime \end{pmatrix}, &
\upsilon^\prime_{abcd} \upsilon^{abcd} &= \gram_4^2 \begin{pmatrix}\vec4 p_a, \vec4 p_b, \vec4 p_c, \vec4 p_d \\
\vec4 p_a^\prime, \vec4 p_b^\prime, \vec4 p_c^\prime, \vec4 p_d^\prime \end{pmatrix} .
\end{align}
\end{subequations}
The contraction of two identical tensor densities \eqref{eq:tensor densities} results in \CM determinants \eqref{eq:CM determinant} of dimension zero to four
\begin{subequations}
\begin{align}
\upsilon_{\mu\nu\rho\sigma} \upsilon^{\mu\nu\rho\sigma} &= \volume^2_0 , &
\upsilon_{a\nu\rho\sigma} \upsilon^{a\nu\rho\sigma} &= \volume_1^2(m_a) , &
\upsilon_{abc\sigma} \upsilon^{abc\sigma} &= \volume_3^2(m_{abc}) , \\
& &
\upsilon_{ab\rho\sigma} \upsilon^{ab\rho\sigma} &= \volume_2^2(m_{ab}) , &
\upsilon_{abcd} \upsilon^{abcd} &= \volume_4^2(m_{abcd}) .
\end{align}
\end{subequations}
The relations between the contractions of the \LC tensor densities with the Gram and \CM determinants suggests their appearance as invariants in calculations in \MST.
The fact that the largest \CM determinant that can be constructed from the \LC tensor density is of dimension four reflects the fact that higher dimensional \CM determinants must vanish in such calculations.

\subsection{Volumes of parallelotopes and simplices}

The volume $\volume_n(m_a, \dots, m_n)$ of an $n$-dimensional parallelotope $\parallel_n(m_a, \dots, m_n)$ with edges~$m_i$ and diagonal $m_{a\cdots n}$ is given by the square root of the \CM determinant \eqref{eq:CM determinant}.
Furthermore, an $n$-dimensional parallelotope is triangulated by $n!$ simplices $\simplex_n(m_a, \dots, m_n)$ with volumes $\flatfrac{\volume_n(m_a, \dots, m_n)}{n!}$.
These simplices are defined by the edges $m_i$, $m_{a\cdots n}$, and additional invariant masses appearing in the \CM determinant \eqref{eq:CM determinant}.
The volumes of the zero and one dimensional parallelotopes are
\begin{align} \label{eq:trivial volumes}
\volume_0 &= \sqrt \eta, &
\volume_1(m_a) = m_a .
\end{align}
Due to the sign in the definition \eqref{eq:Gram determinant vector} the Minkowskian volumes differ by a factor of $\i^{n+1}$ from Euclidean volumes.
Consequently the zero dimensional volume in \eqref{eq:trivial volumes} is imaginary in \MST.
Additionally, this difference affects the area of the two-dimensional parallelogram $\parallel_2(m_{ab})$ and the volume of the four-dimensional parallelotope $\parallel_4(m_{abcd})$.
Since the triangle inequality in \MST is inverted in comparison to Euclidean space \cite{Manoukian:1993}
\begin{equation} \label{eq:triangle inequality}
\begin{array}{rl}
m_a + m_b \geq m_{ab} & \text{for Euclidean space}, \\
m_a + m_b \leq m_{ab} & \text{for \MSTlong}.
\end{array}
\end{equation}
Therefore, the area of the triangle $\triangle_2(m_{ab}; m_a, m_b)$ is defined to be real in \MST as well as in Euclidean space.
In the following we illustrate parallelotopes and simplices in \MST using depictions in Euclidean space, keeping in mind that not all properties are represented adequately.
In particular the triangle inequality \eqref{eq:triangle inequality} cannot be depicted correctly.

\subsection{Determinant identities}

Since invariants in \MST can be identified with volumes that are calculated using determinants, relations between these invariants can be constructed from determinant identities.
On the one hand, specifying the Laplace expansion of the Gram determinant \eqref{eq:Laplace expansion Gram determinant} to the symmetric case allows to express the $n$-dimensional \CM determinant as function of an $n-1$-dimensional \CM determinant and a sum of Gram determinant pairs
\begin{equation} \label{eq:Laplace expansion CM determinant}
\eta \volume^2_n(m_a, \dots, m_n) = m_a^2 \volume^2_{n-1}(m_a, \dots, m_{n-1}) - \sum_{i=b}^n \gram_1^2(m_a, m_i) \gram^2_{n-1}(m_a,m_i;\dots) .
\end{equation}
On the other hand, applying the \DJ determinant identity
\footnote{
The \DJ determinant identity is the two dimensional version of the Sylvester's determinant identity and reads
$
\det M \det M_{i,j}^{i,j} = \begin{vsmallmatrix} \det M_i^i & \det M_i^j \\
\det M_j^i & \det M_j^j \end{vsmallmatrix}
$
where $M_i^j$ corresponds to the submatrix of the matrix $M$ that has the $i$-th row and $j$-th column removed.
}
to the Gram determinant \eqref{eq:Gram determinant} results in
\begin{multline} \label{eq:two-body Gram determinant DJ identity}
\gram^2_{n-1}(m_a, m_b; m_c \dots, m_n) \volume^2_{n-3} (m_d, \dots, m_n) = \\
\volume^2_{n-2}(m_c, \dots, m_n) \gram^2_{n-2}(m_a, m_b; m_d \dots, m_n) \\
- \gram^2_{n-2}(m_a, m_c; m_d, \dots, m_n) \gram^2_{n-2}(m_b, m_c; m_d, \dots, m_n) ,
\end{multline}
and applying it to the symmetric Gram determinant \eqref{eq:symmetric Gram determinant} gives
\begin{multline} \label{eq:CM determinant DJ identity matrix}
\volume^2_n(m_a, \dots, m_n) \volume^2_{n-2} (m_c, \dots, m_n) = \\
\begin{vmatrix*}[l]
\volume^2_{n-1}(m_b, m_c, \dots, m_n) & \gram^2_{n-1}(m_a, m_b; m_c, \dots, m_n) \\
\gram^2_{n-1}(m_a, m_b; m_c, \dots, m_n) & \volume^2_{n-1}(m_a, m_c, \dots, m_n)
\end{vmatrix*}.
\end{multline}
This determinant can either be expressed as
\begin{multline} \label{eq:CM determinant DJ identity}
\volume^2_n(m_a, \dots, m_n) \volume^2_{n-2} (m_c, \dots, m_n) = \\
\volume^2_{n-1}(m_a, m_c, \dots, m_n) \volume^2_{n-1}(m_b, m_c, \dots, m_n) - \gram_{n-1}^4(m_a, m_b; m_c, \dots, m_n) ,
\end{multline}
or as a two dimensional \CM determinant of further \CM determinants
\begin{multline} \label{eq:CM determinant DJ identity square}
\volume^2_n(m_a, m_b , m_c, \dots, m_n) \volume^2_{n-2}(m_c, \dots, m_n) = \\
- \volume_2^2\left(\volume_{n-1}(\m_{ab}, m_c, \dots, m_n); \volume_{n-1}(m_a, m_c, \dots, m_n), \volume_{n-1}(m_b, m_c, \dots, m_n)\right).
\end{multline}
The properties of the two dimensional \CM determinant $\volume_2^2$ will be discussed in more detail in \cref{sec:triangles}.
A symmetric variant of this expression reads
\begin{multline} \label{eq:CM determinant DJ identity square symmetric}
- \volume^2_n(m_a, m_b , m_c, m_d, \dots, m_n) \volume^2_{n-2}(\m_{abc}, m_d, \dots, m_n) = \\
\volume_2^2\left(\volume_{n-1}(\m_{ab}, m_c, m_d, \dots, m_n); \volume_{n-1}(\m_{ac}, m_b, m_d, \dots, m_n), \volume_{n-1}(\m_{bc}, m_a, m_d, \dots, m_n)\right) .
\end{multline}
The non-canonical invariant mass $\m_{ac}$ appearing here can be expressed in terms of canonical invariant masses using momentum conservation \eqref{eq:three-body invariant mass}.

\subsection{Angles of simplices and between particles}

An $n$-dimensional simplex $\simplex_n(m_a, \dots, m_n)$ has $n+1$ vertices $\vertex_{a\cdots n}^a(\m_{ab}, \dots, \m_{a\cdots {n-1}})$.
A characteristic quantity attached to such a vertex is the angle $\alpha_{a\cdots n}^a(\m_{ab}, \dots, \m_{a\cdots n-1})$ between the two $n-1$-dimensional simplices $\simplex_{n-1}(m_a, \m_{ab}, \dots, \m_{a\cdots n-1})$ and $\simplex_{n-1}(m_{a\cdots n}, \m_{ab}, \dots, \m_{a\cdots n-1})$.
Using the \DJ determinant identity \eqref{eq:CM determinant DJ identity} the cosine of this angle can be calculated to be \cite{Byckling:1971vca}
\begin{multline} \label{eq:cosine}
\cos \alpha_{a\cdots n}^a(\m_{ab}, \dots, \m_{a\cdots n-1}) = \\
\frac{\gram^2_{n-1}(m_{a\cdots n}, m_a; \m_{ab}, \dots, \m_{a\cdots n-1})}{\volume_{n-1}(m_{a\cdots n}, \m_{ab}, \dots, \m_{a\cdots n-1})\volume_{n-1}(m_a, \m_{ab}, \dots, \m_{a\cdots n-1})},
\end{multline}
while the sine is given by \cite{Schlafli1950-1}
\begin{multline} \label{eq:sine}
\sin \alpha_{a\cdots n}^a(\m_{ab}, \dots, \m_{a\cdots n-1}) = \\
\frac{\volume_{n-2}(\m_{ab}, \dots, \m_{a\cdots n-1}) \volume_n(m_a, \dots, m_n)}{\volume_{n-1}(m_{a\cdots n}, \m_{ab}, \dots, \m_{a\cdots n-1})\volume_{n-1}(m_a, \m_{ab}, \dots, \m_{a\cdots n-1})} .
\end{multline}
While the cosine \eqref{eq:cosine} can become negative the sine \eqref{eq:sine} is constrained to be positive, since it consists of a product of volumes.
Therefore, the angles that can be described by this pair of equations are constrained to lie within
\begin{equation} \label{eq:angle domain}
0 \leq \alpha_{a\cdots n}^a(\m_{ab}, \dots, \m_{a\cdots n-1}) \leq \pi .
\end{equation}
In \MST these angles can be related to different observables of the involved particles depending on the dimension of the simplex \cite{Byckling:1971vca}.

\section{Two-body interactions} \label{sec:two-body interaction}

Four-momentum conservation ensures that in an interaction of two particles with momenta $\vec4 p_a$ and $\vec4 p_b$ the third particle has momentum
\begin{equation} \label{eq:two-body momentum conservation}
\vec4 p_{ab} := \vec4 p_a + \vec4 p_b .
\end{equation}
and its invariant mass square \eqref{eq:invariant mass} is identical to the mass of the third particle
\begin{align} \label{eq:two-body invariant mass}
\zeta \vec4 p_{ab}^2 &=: \m_{ab}^2 = m_{ab}^2, &
m_{ab}^2 &= m_a^2 + m_b^2 + 2 \zeta\pair{p_a}{p_b} .
\end{align}
Consequently, the two-body \PS is parameterised by a parallelogram $\parallel_2(m_{ab}; m_a, m_b)$ with edges $m_a$ and $m_b$.
Its diagonal is $m_{ab}$ and it is triangulated by two triangles $\simplex_2(m_{ab}; m_a, m_b)$ that are formed by the edges $m_a$, $m_b$, and $m_{ab}$.
See \cref{fig:parallelogram} for a depiction.

\begin{figure}
\begin{panels}{.4}
\includepgf*{parallelogram}
\caption{\PS parallelogram.} \label{fig:parallelogram}
\panel{.6}
\includepgf*{two-body}
\caption{\PS diagram with frame dependent quantities.} \label{fig:triangle}
\end{panels}
\caption[Diagrams of the two-body \PSlong]{
Two-body \PS parallelogram $\parallel_2(m_{ab}; m_a, m_b)$ in panel \subref{fig:parallelogram} and the \PS diagram given by the triangle $\simplex_2(m_a, m_b)$ in panel \subref{fig:triangle}.
In panel \subref{fig:triangle} the energy $\energy_a(m_{ab})$ (purple dash-dotted projection) and the absolute values of the momentum $\momentum_a(m_{ab})$ (green dotted height) in the rest frame of the particle with mass $m_{ab}$ as well as the relative rapidity $\zeta_{ab}^a$ between these two frames are also indicated, \cf \cref{sec:rapidity,sec:energy and momentum}.
Note that despite this Euclidean depiction the opposite Minkowskian triangle inequality \eqref{eq:triangle inequality} must be satisfied in reality.
} \label{fig:two-body PS}
\end{figure}

\subsection{Parallelograms and triangles} \label{sec:triangles}

For two-body processes the Gram determinants \eqref{eq:Gram determinant} have mass-dimension two.
Using the expansion \eqref{eq:two-body invariant mass} the Gram determinant at the vertex $\vertex_b^a$ reads
\begin{equation} \label{eq:two-body Gram}
\gram_1^2(m_a, m_b) =
\zeta\gram_1^2 \begin{pmatrix}\vec4 p_a \\
\vec4 p_b \end{pmatrix} =
\zeta\pair{p_a}{p_b} = \frac{m_{ab}^2 - m_a^2 - m_b^2}{2} .
\end{equation}
Four-momentum conservation \eqref{eq:two-body momentum conservation} ensures that two Gram determinants can be added to result in a squared mass
\begin{align} \label{eq:two-body Gram sum}
\gram_1^2(m_a, m_{ab}) - \gram_1^2(m_a, m_b) &= m_a^2, &
\gram_1^2(m_{ab}, m_a) + \gram_1^2(m_{ab}, m_b) &= m_{ab}^2,
\end{align}
The two-body \CM determinant \eqref{eq:CM determinant} has mass dimension four and reads
\begin{equation} \label{eq:area}
\volume_2^2(m_{ab}; m_a, m_b) = - \frac{\eta}4 \begin{vmatrix*}[l]
\header0 & \header1 & \header1 & \header1 \\
\header1 & \header0 & m_a^2 & m_{ab}^2 \\
\header1 & m_a^2 & \header0 & m_b^2 \\
\header1 & m_{ab}^2 & m_b^2 & \header0
\end{vmatrix*} .
\end{equation}
Its square root corresponds the area of the parallelogram $\parallel_2(m_{ab}; m_a, m_b)$ shown in \cref{fig:parallelogram}.
The area can also be written using Heron's formula \cite{Hero:orig}
\begin{equation} \label{eq:Heron}
\volume_2^2(m_{ab}; m_a, m_b) = 4 \eta \func t(m_a; m_a, m_b) \func t(m_b; m_a, m_b) \func t(m_{ab}; m_a, m_b) \func s(m_{ab}; m_a, m_b) ,
\end{equation}
where the $t$~function and the semiperimeter $s$ of the triangle are
\begin{align}
\func t(m_i; m_a, m_b) &:= \func s(m_{ab}; m_a, m_b) - m_i , &
\func s(m_{ab}; m_a, m_b) &:= \frac{m_{ab} + m_a + m_b}2 .
\end{align}
The corresponding triangle $\triangle_2(m_{ab}; m_a, m_b)$ depicted in \cref{fig:triangle} has the area $\volume_2(m_{ab}; m_a, m_b) / 2$.
The sign $\eta$ defined in \eqref{eq:metric determinant} ensures that it is real for the Euclidean and the Minkowskian triangle inequality \eqref{eq:triangle inequality}.
Relevant limiting cases for the area of the parallelogram in \MST are
\begin{equation} \label{eq:two-body limits}
\volume_2(m_{ab}; m_a, m_b) = \frac{\sqrt{-\eta}}2
\begin{cases}
m_{ab} \sqrt{m_{ab}^2 - 4m_a^2} & \text{for } m_b = m_a, \\
m_{ab}^2 - m_a^2 & \text{for } m_b = 0, \\
m_{ab}^2 & \text{for } m_b = m_a = 0, \\
0 & \text{for } m_b = m_{ab} - m_a .
\end{cases}
\end{equation}
Hence, the area of the parallelogram $\parallel_2(m_{ab}; m_a, m_b)$ vanishes when the triangle inequality \eqref{eq:triangle inequality} is saturated.
Therefore, the physical regime of the two-body interaction \eqref{eq:two-body momentum conservation} is characterised by a positive area for this parallelogram.
Furthermore, the area can be connected to the Gram determinant \eqref{eq:two-body Gram}.
On the one hand, specifying the \DJ determinant identity \eqref{eq:CM determinant DJ identity} to two dimensions results in
\begin{equation} \label{eq:two-body DJ identity}
\eta \volume_2^2(m_a,m_b) = m_a^2 m_b^2 - \gram_1^4(m_a,m_b) .
\end{equation}
On the other hand specifying the Laplace expansion \eqref{eq:Laplace expansion CM determinant} and employing four-momentum conservation \eqref{eq:two-body Gram sum} results in
\begin{equation} \label{eq:two-body Laplace}
\eta \volume_2^2(m_a, m_b) = \gram_1^2(m_a, m_{ab}) \gram_1^2(m_b, m_{ab}) - m_{ab}^2 \gram_1^2(m_a, m_b) .
\end{equation}

\subsection{Rapidity, Lorentz factor, and velocity norm} \label{sec:rapidity}

In \MST the angle at the vertex $\vertex_{ab}^a$ of the triangle $\triangle_2(m_{ab}; m_a, m_b)$ corresponds to the rapidity~$\zeta_{ab}^a$ between the rest frames of the particles with masses $m_{ab}$ and $m_a$, see \cref{fig:triangle} for a depiction.
The (co-)sine of this angle can be determined by specifying the generic equations \eqref{eq:sine,eq:cosine} to two dimensions
\begin{align}
\cosh\zeta_{ab}^a &= \cos \i\zeta_{ab}^a = \frac{\gram_1^2(m_a, m_{ab})}{\volume_1(m_a)\volume_1(m_{ab})}, &
\i\sinh\zeta_{ab}^a &= \sin \i\zeta_{ab}^a = \frac{\volume_0 \volume_2(m_{ab})}{\volume_1(m_a)\volume_1(m_{ab})},
\end{align}
here the trigonometric functions are hyperbolic since the volume of the zero dimensional parallelotope \eqref{eq:trivial volumes} is imaginary.
Therefore, the rapidity can be expressed as
\begin{align} \label{eq:rapidity}
\cosh \zeta_{ab}^a &= \frac{\gram_1^2(m_a, m_{ab})}{m_a m_{ab}}, &
\tanh \zeta_{ab}^a &= \frac{\volume_2(m_{ab})}{\gram_1^2(m_a, m_{ab})}, &
\sinh \zeta_{ab}^a &= \frac{\volume_2(m_{ab})}{m_a m_{ab}} .
\end{align}
Where the two-body Gram determinant and the area of the parallelogram are given by \eqref{eq:two-body Gram} and \eqref{eq:area}, respectively.
The last of these equations constitutes the Minkowskian equivalent to the Euclidean relation that connects the area of a triangle to one of its angles and the lengths of the two adjacent edges.
The rapidity is symmetric under exchange of indices and the sum of the angles connected to the longest side is equal to the third angle
\begin{align} \label{eq:rapidity symmetry}
\zeta_a^{ab} &= \zeta_{ab}^a, &
\zeta_a^b &= \zeta_a^{ab} + \zeta_{ab}^b .
\end{align}
The hyperbolic trigonometric functions of the rapidity appearing in \eqref{eq:rapidity} can be identified with the relative Lorentz factor~$\gamma_{ab}^a$ between the two rest frames of the particles with masses $m_{ab}$ and $m_a$, the corresponding absolute value of the relative velocity $v_{ab}^a := \abs{\vec3 v_{ab}^a}$, and their product
\begin{align} \label{eq:Lorentz velocity}
\gamma_{ab}^a &:= \cosh \zeta_{ab}^a, &
v_{ab}^a &:= \tanh \zeta_{ab}^a, &
w_{ab}^a := \gamma_{ab}^a v_{ab}^a &= \sinh \zeta_{ab}^a,
\end{align}
Hence they are given by
\begin{align} \label{eq:Lorentz velocity identity}
\gamma_{ab}^a &= \frac{\gram_1^2(m_a, m_{ab})}{m_a m_{ab}}, &
v_{ab}^a &= \frac{\volume_2(m_{ab})}{\gram_1^2(m_a, m_{ab})}, &
w_{ab}^a &= \frac{\volume_2(m_{ab})}{m_a m_{ab}} .
\end{align}
These quantities have the same symmetry as the rapidity \eqref{eq:rapidity symmetry} and obey the addition formulas
\begin{align}
\gamma_a^b &= \gamma_a^{ab} \gamma_{ab}^b (1 + v_a^{ab} v_{ab}^b), &
v_a^b &= \frac{v_a^{ab} + v_{ab}^b}{1 + v_a^{ab} v_{ab}^b}, &
w_a^b &= \gamma_a^{ab} \gamma_{ab}^b (v_a^{ab} + v_{ab}^b) .
\end{align}
These relation can be derived from the identities \eqref{eq:two-body DJ identity,eq:two-body Laplace}.
Using the first of the identities \eqref{eq:Lorentz velocity identity} the product of two four-momenta appearing in \eqref{eq:two-body invariant mass} can be parameterised in terms of the two corresponding masses and the Lorentz factor between their rest frames
\begin{equation} \label{eq:Gram rapidity}
\zeta\pair{p_a}{p_b} = m_a m_b \gamma_a^b .
\end{equation}
Therefore, relation \eqref{eq:two-body invariant mass} can be interpreted as the law of cosines for the triangle $\triangle_2(m_{ab}; m_a, m_b)$.
The corresponding law of sines is related to the invariant
\begin{equation}
2 R(m_{ab}; m_a, m_b) := \frac{m_{ab} m_a m_b}{\volume_2(m_{ab}; m_a, m_b)} = \frac{m_{ab}}{w_a^b} = \frac{m_a}{w_{ab}^b} = \frac{m_b}{w_{ab}^a} ,
\end{equation}
that can be identified with the radius of the circumcircle around the triangle.

\subsection{Energy and three-momentum norm} \label{sec:energy and momentum}

The energy and the absolute value of the three-momentum $\momentum_a(m_{ab}) := \abs{\vec3 \momentum_a(m_{ab})}$ of the particle with mass $m_a$ in the rest frame of particle with mass $m_{ab}$ are expressed via the Lorentz factor and the absolute value of the velocity between these two rest frames \eqref{eq:Lorentz velocity}
\begin{align} \label{eq:energy and momentum definition}
\energy_a(m_{ab}) &:= m_a \gamma_{ab}^a , &
\momentum_a(m_{ab}) &:= m_a w_{ab}^a .
\end{align}
Therefore, energies and momenta can be interpreted as projections and heights within the triangle $\triangle_2(m_{ab}; m_a, m_b)$, as depicted in \cref{fig:triangle}.
Using the identities \eqref{eq:Lorentz velocity identity} they are given by
\begin{align} \label{eq:energy and momentum identities}
\energy_a(m_{ab}) &= \frac{\gram_1^2(m_a, m_{ab})}{m_{ab}}, &
\momentum_a(m_{ab}) &= \frac{\volume_2(m_{ab})}{m_{ab}} .
\end{align}
Consequently, the identities
\begin{subequations} \label{eq:energy momentum special cases}
\begin{align}
\energy_a(m_a) &= m_a, &
m_b \energy_a(m_b) &= m_a \energy_b(m_a), & \\
\momentum_a(m_a) &= 0, &
m_b \momentum_a(m_b) &= m_a \momentum_b(m_a), &
\momentum_a(m_{ab}) &= \momentum_b(m_{ab}),
\end{align}
\end{subequations}
can be directly extracted from the definitions \eqref{eq:energy and momentum identities} and \cref{fig:triangle}.
Expressing the \DJ determinant identity \eqref{eq:two-body DJ identity} in terms of energy and momentum shows that the Pythagorean theorem corresponds to the energy-momentum relation
\begin{equation} \label{eq:energy-momentum relation}
m_a^2 = \energy_a^2(m_{ab}) + \eta \momentum_a^2(m_{ab}) .
\end{equation}
The momentum product appearing in \eqref{eq:two-body invariant mass} can also be formulated frame-dependently using these quantities together with the opening angle $\theta_a^b(m_{ab})$ between the particles with mass $m_a$ and $m_b$ in the rest frame of particle with mass $m_{ab}$
\begin{equation} \label{eq:Gram angle}
\zeta\pair{p_a}{p_b} = \energy_a(m_{ab}) \energy_b(m_{ab}) - \momentum_a(m_{ab}) \momentum_b(m_{ab}) \cos \theta_a^b(m_{ab}) .
\end{equation}
The comparison with the parameterisation \eqref{eq:Gram rapidity} shows that in the rest frame of the particle with mass $m_{ab}$ the particles with masses $m_a$ and $m_b$ are back-to-back
\begin{equation} \label{eq:two-body angle}
\cos \theta_a^b(m_{ab}) = \frac{\energy_a(m_{ab}) \energy_b(m_{ab}) - m_a m_b \gamma_a^b}{\momentum_a(m_{ab}) \momentum_b(m_{ab})} = -1 .
\end{equation}
This expression can be interpreted as a degenerate case of the spherical law of cosines \eqref{eq:three-body cosine physical} introduced in \cref{sec:opening angle}.

\subsection{On-shell four-momentum and solid angle differential} \label{sec:solid angle differential}

The $n$-body \PS \eqref{eq:momentum PS} depends on $n$ four-momentum differentials that are constrained to be on-shell.
In order to express the constrained four-momentum differential for $\vec4 p_a$ when it dependents on a reference frame another particle with mass $m_b$ has to be introduced.
In the rest frame of their invariant mass $\m_{ab}$ the constrained four-momentum differential reads
\begin{multline} \label{eq:constrained four-momentum differential}
\int \diractwopi(\zeta \vec4 p_a^2 - m_a^2) \dtwopi^4 \vec4 p_a = \\
\int \diractwopi(\energy_a^2(\m_{ab}) + \eta \vec3 p_a^2(\m_{ab}) - m_a^2) \dtwopi E_a(\m_{ab}) \dtwopi^3 \vec3 p_a(\m_{ab}) =
\frac{\dtwopi^3 \vec3 p_a(\m_{ab})}{2 \energy_a(\m_{ab})} .
\end{multline}
By extending the reference frame with two more particles, as depicted in \cref{fig:solid angle definition}, the components of the three-momentum can be expressed using angular variables
\begin{subequations}
\begin{align}
\momentum_a^x(\m_{ab};\m_{abc};\m_{abcd}) &= \momentum_a(\m_{ab}) \sin \theta_{abc}^a(\m_{ab}) \cos \phi_{abcd}^a(\m_{ab}, \m_{abc}), \\
\momentum_a^y(\m_{ab};\m_{abc};\m_{abcd}) &= \momentum_a(\m_{ab}) \sin \theta_{abc}^a(\m_{ab}) \sin \phi_{abcd}^a(\m_{ab}, \m_{abc}), \\
\momentum_a^z(\m_{ab};\m_{abc};\m_{abcd}) &= \momentum_a(\m_{ab}) \cos \theta_{abc}^a(\m_{ab}) .
\end{align}
\end{subequations}
In the rest frame of $\m_{ab}$ the polar angle $\theta_{abc}^a(\m_{ab})$ measures the opening angle between the pair $\vec3 p_a$ and $\vec3 p_b$ and the pair $\vec3 p_c$ and $\vec3 p_{abc}$, both of which are back-to-back \eqref{eq:two-body angle}.
Since the polar opening angle is a property of the three-body system it will be discussed in \cref{sec:opening angle}.
The definition of the azimuthal angle requires the introduction of a further particle to the reference frame.
The angle between the two planes defined in \cref{fig:solid angle definition} is given by the azimuthal angle $\phi_{abcd}^a(\m_{ab}, \m_{abc})$.
Since the azimuthal decay plane angle appears in a four-body system it will be discussed in \cref{sec:decay plane angle}.

\begin{figure}
\begin{panels}{.35}
\includepgf*{planes-3}
\caption{Decay planes.} \label{fig:decay planes}
\panel{.28}
\includepgf*{vectors}
\caption{Coordinate system.}
\label{fig:spherical coordinate system}
\panel{.37}
\includepgf*{solid-angle}
\caption{\PS diagram.} \label{fig:invariant solid angle}
\end{panels}
\caption[Depiction of the solid angle differential]{
Angular variables appearing in a four-body process in the rest frame of the invariant mass $\m_{ab}$ in panel \subref{fig:decay planes}, resulting spherical coordinate system in panel \subref{fig:spherical coordinate system}, and the corresponding invariant description in panel \subref{fig:invariant solid angle}.
In the rest frame of the invariant mass $\m_{ab}$ the particles with masses $m_a$ and $m_b$ are back-to-back \eqref{eq:two-body angle}.
Only after the introduction of another particle with mass $m_c$ that is itself back-to-back with $\m_{abc}$ it is possible to define a polar opening angle $\theta_{abc}^a(\m_{ab})$ between these pairs \eqref{eq:three-body cosine physical}.
In order to define an azimuthal angle $\phi_{abcd}^a(\m_{ab}, \m_{abc})$ between two planes \eqref{eq:four-body cosine} a further particle with mass $m_d$ is necessary.
Together these angles define the solid angle of the two-body process of $m_a$ and $m_b$ with respect to the reference frame defined by $m_c$ and $m_d$.
} \label{fig:solid angle definition}
\end{figure}

Therefore, the three-momentum differential appearing in \eqref{eq:constrained four-momentum differential} can be written as
\footnote{
Note the transition from the normalisation \eqref{eq:barred normalisation} to the normalisation \eqref{eq:tilded normalisation}.
}
\begin{equation}
\begin{split}
\dtwopi^3 \vec3 \momentum_a(\m_{ab}) &= 4 \momentum^2(\m_{ab}) \dtwopi p_a(\m_{ab}) \dfourpi^2 \Omega_{abcd}^a(\m_{ab}; \m_{abc}) \\
&= 4 \energy_a (\m_{ab}) \momentum_a(\m_{ab}) \dtwopi E_a(\m_{ab}) \dfourpi^2 \Omega_{abcd}^a(\m_{ab}; \m_{abc})
,
\end{split}
\end{equation}
where the solid angle differential is defined as
\begin{equation} \label{eq:solid angle differential}
\dfourpi^2 \Omega_{abcd}^a(\m_{ab}; \m_{abc}) = \sin \theta_{abc}^a(\m_{ab}) \dfourpi \theta_{abc}^a(\m_{ab}) \dfourpi \phi_{abcd}^a(\m_{ab}, \m_{abc}) .
\end{equation}
with
\footnote{
In the following we will neglect the sign that appears when the differential of a cosine is introduced and assume that it is absorbed in a swap of the integration limits.
}
\begin{equation} \label{eq:polar angle differential}
\sin \theta_{abc}^a(\m_{ab}) \dfourpi \theta_{abc}^a(\m_{ab}) = - \dfourpi \cos \theta_{abc}^a(\m_{ab}) .
\end{equation}
These angles are defined over the domain
\begin{align} \label{eq:solid angle domain}
-1 &\leq \cos \theta_{abc}^a(\m_{ab}) \leq 1, &
0 &\leq \phi_{abcd}^a(\m_{ab}, \m_{abc}) \leq 2 \pi.
\end{align}
Due to the normalisation \eqref{eq:tilded normalisation} they integrate to
\begin{align} \label{eq:solid angle differential integrated}
\int \dfourpi \cos \theta_{abc}^a(\m_{ab}) &= \frac1{2 \pi}, &
\int \dfourpi \phi_{abcd}^a(\m_{ab}, \m_{abc}) &= \frac1{2}, &
\iint \dfourpi^2 \Omega_{abcd}^a(\m_{ab}; \m_{abc}) &= \frac1{4 \pi} .
\end{align}
The solid angle differential is depicted in two different parameterisations in \cref{fig:solid angle}.
Using the frame independent formulation of the energy and the absolute value of the three-momentum \eqref{eq:energy and momentum identities} the constrained four-momentum differential \eqref{eq:constrained four-momentum differential} can therefore be expressed as
\begin{equation} \label{eq:four-momentum differential}
\int \diractwopi(\zeta \vec4 p_a^2 - m_a^2) \dtwopi^4 \vec4 p_a = 2 \frac{\volume_2(\m_{ab})}{\m_{ab}^2} \dtwopi G_1^2(m_a, \m_{ab}) \dfourpi^2 \Omega_{abcd}^a(\m_{ab}; \m_{abc}) .
\end{equation}

\subsection{Two-body \PSlong} \label{sec:two-body PS}

Specifying the generic definition of the differential \PS \eqref{eq:momentum PS} to two particles
\begin{equation}
\d^2 \Phi_2(\vec4 p_{ab}; \vec4 p_a, \vec4 p_b) = \zeta \vec4 p_{ab}^2 \iiiint \diractwopi^4(\vec4 p_{ab} - \vec4 p_a - \vec4 p_b) \int \diractwopi(\zeta \vec4 p_b^2 - m_b^2) \dtwopi^4 \vec4 p_b \int \diractwopi(\zeta \vec4 p_a^2 - m_a^2) \dtwopi^4 \vec4 p_a,
\end{equation}
and using the four-dimensional Dirac distribution to integrate over the four-momentum of the second particle results in
\begin{equation}
\d^2 \Phi_2(\vec4 p_{ab}; \vec4 p_a, m_b) = \zeta \vec4 p_{ab}^2 \int \diractwopi\left(\zeta (\vec4 p_{ab} - \vec4 p_a)^2 - m_b^2\right) \int \diractwopi(\zeta \vec4 p_a^2 - m_a^2) \dtwopi^4 \vec4 p_a .
\end{equation}
Using the constrained four-momentum differential \eqref{eq:four-momentum differential} of the first particle the differential \PS
reads
\begin{multline}
\d^2 \Phi_2(m_{ab}; m_a, m_b; m_c, m_d) = \volume_2(\m_{ab}) \dfourpi^2 \Omega_{abcd}^a(m_{ab}; \m_{abc}) \\
2 \int \diractwopi\left(m_{ab}^2 - 2 \gram_1^2(m_a, m_{ab}) + m_a^2 - m_b^2\right) \dtwopi G_1^2(m_a, m_{ab}) .
\end{multline}
Finally, the integral over the Gram determinant results for the differential \PS in
\begin{equation} \label{eq:differential two-body PS}
\d^2 \Phi_2(m_{ab}; m_c, m_d) = \volume_2(m_{ab}) \dfourpi^2 \Omega_{abcd}^a(m_{ab}; \m_{abc}) .
\end{equation}
After integrating over the external solid angle differential \eqref{eq:solid angle differential integrated}, the \PS is constant and frame-independent
\begin{equation} \label{eq:integrated two-body PS}
\Phi_2(m_{ab}) = \frac{\volume_2(m_{ab})}{4 \pi} .
\end{equation}
Hence the \PS of the interactions between a particle with mass $m_{ab}$ and two particles with masses $m_a$ and $m_b$ is governed by the area of the parallelogram $\parallel_2(m_{ab}; m_a, m_b)$ depicted in \cref{fig:parallelogram}.
The trivial \PS diagram of the two-body \PS is depicted in \cref{fig:triangle}.

\subsection{Decays into two final particles}

After the integration over the solid angle differential the two-body \PS \eqref{eq:integrated two-body PS} leads for the decay width \eqref{eq:decay width} of a particle with mass $m_{ab}$ that decays into two particles with masses $m_a$ and $m_b$ in its own rest frame to
\begin{equation}
\Gamma_2(m_{ab}; m_a, m_b) = \frac{\abs{\mathcal A_3}^2}{8\pi} \frac{\volume_2(m_{ab})}{m_{ab}^3} .
\end{equation}
With the Lagrangian and amplitude of the trivial interaction of three scalar particles
\begin{align}
\mathcal L_3 &\supset \frac{g}{3!} \phi_{ab} \phi_a \phi_b, &
\mathcal A_3 &= \i g,
\end{align}
where the coupling constant $g$ has mass dimension one,
the decay width reaches for $m_a = m_b = 0$ the maximal value of
\begin{equation} \label{eq:maximal two-body PS}
\Gamma_2^{\max}(m_{ab}) = \frac{g^2}{16 \pi m_{ab}} .
\end{equation}

\section{Three-body interactions} \label{sec:three-body interaction}

The invariant mass \eqref{eq:invariant mass} of three particles
\begin{align}
m_{abc}^2 = \m_{abc}^2 &:= \zeta \vec4 p_{abc}^2 , &
\vec4 p_{abc} &:= \vec4 p_a + \vec4 p_b + \vec4 p_c,
\end{align}
can be expressed in terms of the invariant masses of two particles \eqref{eq:two-body invariant mass} via
\begin{equation} \label{eq:three-body invariant mass}
m_{abc}^2 = \m_{ab}^2 + \m_{bc}^2 + \m_{ac}^2 - m_a^2 - m_b^2 - m_c^2 ,
\end{equation}
and allows to eliminate the non-canonical invariant mass $\m_{ac}$.
The three-body \PS is governed by the volume $\volume_3(m_{abc}; m_a, m_b, m_c)$ of a three-dimensional parallelepiped $\parallel_3(m_{abc}; m_a, m_b, m_c)$.
It can be triangulated by six tetrahedra $\simplex_3(m_{abc}; m_a, m_b, m_c)$ with volumes $\volume_3 / 6$.
The freedom to eliminate one of the three two-particle invariant masses results in three different triangulations, only one of which we call canonical.
The rapidities, velocities, and Lorentz factors between the different rest frames as well as the energies and momenta in the different rest frames can be derived by calculating the angles, heights, and projections on the different faces of the tetrahedra as demonstrated in \cref{sec:rapidity,sec:energy and momentum}, respectively.

\subsection{Parallelepipeds and tetrahedra}

\begin{figure}
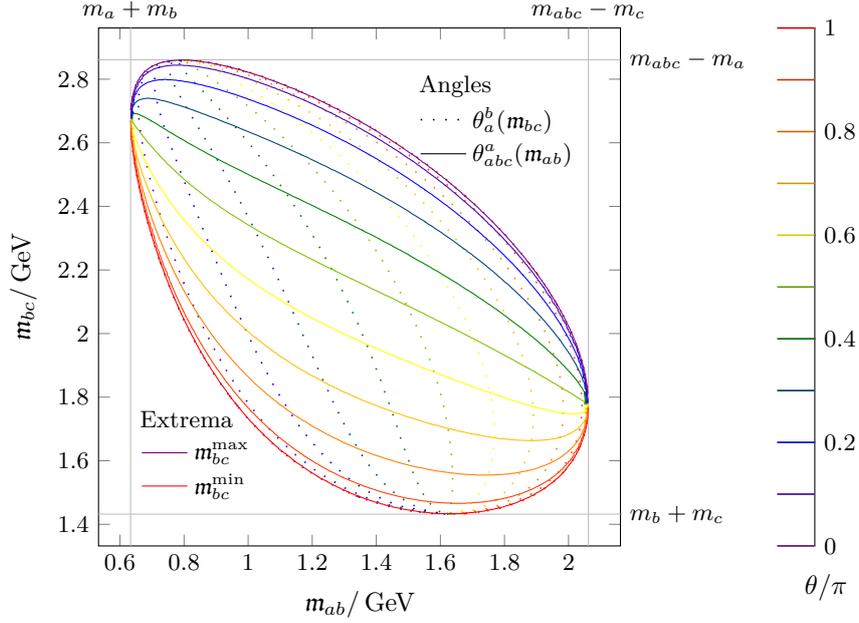

\begin{panels}{.9}
\includepgf{opening-angle}
\end{panels}
\caption[Three-body Dalitz plot]{
Dalitz plot of the integration limits \eqref{eq:three-body integration limits} for the masses $m_a = \unit[139]{MeV}$, $m_b = \unit[494]{MeV}$, $m_c = \unit[938]{MeV}$, $m_{abc} = \unit[3]{GeV}$, corresponding to a resonance of \unit[3]{GeV} that decays into $p \widebar K^0 \pi^+$.
The allowed region is covered by lines of constant polar opening angles \eqref{eq:three-body cosine}.
} \label{fig:Dalitz plot}
\end{figure}

The three-body Gram determinant \eqref{eq:Gram determinant} has mass-dimension four.
At the vertex $\vertex_{abc}^a(\m_{ab})$ between the masses $m_{abc}$ and $m_a$ and the invariant mass $\m_{ab}$ it is given by
\begin{equation}
\gram_2^2(m_a, m_{abc}; \m_{ab}) =
\gram_2^2
\begin{pmatrix*}[r]
\vec4 p_a, \vec4 p_{ab} \\
\vec4 p_{abc}, \vec4 p_{ab}
\end{pmatrix*}
= \eta
\begin{vmatrix*}[r]
\pair{p_a}{p_{abc}} & \pair{p_a}{p_{ab}} \\
\pair{p_{ab}}{p_{abc}} & \pair{p_{ab}}{p_{ab}}
\end{vmatrix*} ,
\end{equation}
and a Laplace expansion in terms of minors \eqref{eq:Laplace expansion Gram determinant} leads to
\begin{equation} \label{eq:three-body Laplace}
\eta \gram_2^2(m_a, m_{abc}; \m_{ab}) = \gram_1^2(m_a, m_{abc}) \m_{ab}^2 - \gram_1^2(m_a, \m_{ab}) \gram_1^2(m_{abc}, \m_{ab}) .
\end{equation}
The three-body \CM determinant \eqref{eq:CM determinant} has mass dimension six and reads
\begin{equation}
\volume_3^2(m_{abc}; m_a, m_b, m_c) = \frac18 \begin{vmatrix*}[l]
\header0 & \header1 & \header1 & \header1 & \header1 \\
\header1 & \header0 & m_a^2 & \m_{ab}^2 & m_{abc}^2 \\
\header1 & m_a^2 & \header0 & m_b^2 & \m_{bc}^2 \\
\header1 & \m_{ab}^2 & m_b^2 & \header0 & m_c^2 \\
\header1 & m_{abc}^2 & \m_{bc}^2 & m_c^2 & \header0 \\
\end{vmatrix*} .
\end{equation}
Limiting cases of the volume of the parallelepiped are
\begin{equation} \label{eq:three-body sine limits}
\volume_3(m_{abc}; m_a, m_b, m_c) = \frac12
\begin{cases}
\m_{ab} \m_{bc} \m_{ac} & \text{for } m_a = m_b = m_c = 0, \\
0 & \text{for } m_{abc} = m_a + m_b + m_c, \\
0 & \text{for } \m_{bc} = \m_{bc}^{\nicefrac{\max}{\min}}(\m_{ab}) .
\end{cases}
\end{equation}
Since the volume vanishes if the masses $m_a$, $m_b$, and $m_c$ add up exactly to the mass $m_{abc}$, the physically regime of the three-body process correspond to positive volumes for the parallelepiped $\parallel_3(m_{abc}; m_a, m_b, m_c)$.
Using the \DJ determinant identity \eqref{eq:CM determinant DJ identity square symmetric} the volume can be compactly written as
\begin{equation}
\volume_3^2(m_a, m_b, m_c) = - \frac{\volume_2^2(\volume_2(\m_{ac}, m_b); \volume_2(\m_{ab}, m_c), \volume_2(\m_{bc}, m_a))}{m_{abc}^2}.
\end{equation}
From the two-body limits \eqref{eq:two-body limits} follows then that the volume of the parallelepiped vanishes when the areas of the sides obey
\begin{equation}
\volume_2(\m_{ac}, m_b) = \volume_2(\m_{ab}, m_c) + \volume_2(\m_{bc}, m_a).
\end{equation}
This condition can be exploited to calculate a further constellation that results in a vanishing volume for any masses that obey $m_a + m_b + m_c < m_{abc}$.
It is reached when the invariant mass $\m_{bc}$ assumes as a function of the invariant mass $\m_{ab}$ one of its extremal values.
The extrema can be observed in the Dalitz plot \cite{Dalitz:1953cp} shown in \cref{fig:Dalitz plot} and are given by
\begin{equation} \label{eq:three-body integration limits}
{\m_{bc}^{\nicefrac{\max}{\min}}}^2(\m_{ab}) = \energy_{bc}^+(\m_{ab})^2 + \eta \momentum_{bc}^\pm(\m_{ab})^2 ,
\end{equation}
where the sum of energies and the sum and difference of momenta in the rest frame of the invariant mass $\m_{ab}$ are given by
\begin{align} \label{eq:three-body integration limits ingredients}
\energy_{bc}^+(\m_{ab}) &= \energy_b(\m_{ab}) + \energy_c(\m_{ab}), &
\momentum_{bc}^\pm(\m_{ab}) &= \momentum_b(\m_{ab}) \pm \eta \momentum_c(\m_{ab}) .
\end{align}
Since the invariant description of the energy \eqref{eq:energy and momentum identities} depends on the Gram determinant \eqref{eq:two-body Gram} the sum in the definition of the energy \eqref{eq:three-body integration limits ingredients} can become a difference when calculating the production cross section if one of the four-momenta corresponds to the incoming particle with swapped momentum.
An Euclidean depiction of these quantities is shown in \cref{fig:three-body PS masses}.
In contrast to Euclidean space, in \MST the momentum difference results in the minimal invariant mass while the momentum sum leads to the maximal invariant mass.
The volume is then given as a function of these extrema as
\begin{equation}
\volume_3^2(m_a, m_b, m_c) = \frac{{\m_{bc}^{\vphantom{i}\max}}^2(\m_{ab}) - \m_{bc}^2}2 \m_{ab}^2 \frac{\m_{bc}^2 - {\m_{bc}^{\min}}^2(\m_{ab})}2 .
\end{equation}
Specifying the Laplace expansion \eqref{eq:Laplace expansion CM determinant} to the three dimensional case shows that the volumes and Gram determinants are connected via
\begin{multline}
\eta \volume_3^2(m_a, m_b, m_c) = \\
m_a^2 \volume_2^2(m_b, m_c) - \gram_1^2(m_a, m_b) \gram_2^2(m_a, m_b; m_c) - \gram_1^2(m_a, m_c) \gram_2^2(m_a, m_c; m_b) .
\end{multline}

\subsection{Opening angle between two particles} \label{sec:opening angle}

The opening angle between the particles with masses $m_a$ and $m_{abc}$ in the rest frame of the invariant mass $\m_{ab}$
\begin{equation}
\cos_{abc}^a(\m_{ab}) := \frac{\vec3 p_a(\m_{ab}) \cdot \vec3 p_{abc}(\m_{ab})}{\abs{\vec3 p_a(\m_{ab})} \abs{\vec3 p_{abc}(\m_{ab})}}.
\end{equation}
can be identified with the dihedral angle of the tetrahedron $\simplex_3(m_{abc}; m_a, m_b, m_c)$ at the edge $\m_{ab}$ when it forms a vertex $\vertex_{abc}^a(\m_{ab})$ with $m_a$ and $m_{abc}$ \cite{Byckling:1969sx}.
Specifying the general expression for the cosine \eqref{eq:cosine} to the three dimensional case results in the spherical law of cosines
\begin{equation} \label{eq:three-body cosine}
\cos \theta_{abc}^a(\m_{ab}) = \frac{\gram_2^2(m_a, m_{abc}; \m_{ab})}{\volume_2(m_a, \m_{ab}) \volume_2(m_{abc}, \m_{ab})} ,
\end{equation}
where the Gram determinant can be expressed by the Laplace expansion \eqref{eq:three-body Laplace}.
This expression can then be solved for the invariant mass
\begin{equation} \label{eq:three-body variable transformation}
\m_{bc}^2 = m_{abc}^2 + m_a^2 -
2 \frac{g_{abc}^a(\m_{ab})}{\m_{ab}^2} ,
\end{equation}
where
\begin{equation}
g_{abc}^a(\m_{ab}) = \eta \volume_2(m_{abc}, \m_{ab}) \volume_2(m_a, \m_{ab}) \cos \theta_{abc}^a(\m_{ab}) + \gram_1^2(m_{abc}, \m_{ab}) \gram_1^2(m_a, \m_{ab}) .
\end{equation}
and the derivative with respect to the polar opening angle reads
\begin{equation} \label{eq:three-body derivative}
\dv[\m_{bc}^2]{\cos \theta_{abc}^a(\m_{ab})} = - 2 \eta \frac{\volume_2(m_{abc}, \m_{ab}) \volume_2(m_a, \m_{ab})}{\m_{ab}^2} .
\end{equation}
Using the Laplace expansion of the Gram determinant \eqref{eq:three-body Laplace} and the invariant expressions for the energy and the momenta \eqref{eq:energy and momentum identities} allows to express the cosine of the opening angle \eqref{eq:three-body cosine} as function of
energies and momenta
\begin{equation} \label{eq:three-body cosine physical}
\cos \theta_{abc}^a(\m_{ab}) = \eta \frac{m_{abc} \energy_a(m_{abc}) - \energy_{abc}(\m_{ab}) \energy_a(\m_{ab})}{\momentum_{abc}(\m_{ab}) \momentum_a(\m_{ab})}.
\end{equation}
Using the relation between the energy and the Lorentz factor \eqref{eq:energy and momentum definition} this expression can be compared to the degenerate case \eqref{eq:two-body angle}.
The sine of the polar opening angle between the particles with masses $m_{abc}$ and $m_a$ in the rest frame of the invariant mass $\m_{ab}$ is given by specifying the general definition \eqref{eq:sine} to three dimensions
\begin{equation} \label{eq:three-body sine}
\sin \theta_{abc}^a(\m_{ab}) = \frac{\m_{ab} \volume_3(m_{abc}; m_a, m_b, m_c)}{\volume_2(m_{abc}, \m_{ab}) \volume_2(m_a, \m_{ab})} .
\end{equation}
Therefore, the tangent of the polar opening angle is independent of the two-dimensional volumes
\begin{equation} \label{eq:three-body tangent}
\tan \theta_{abc}^a(\m_{ab}) = \frac{\m_{ab} \volume_3(m_{abc}; m_a, m_b, m_c)}{\gram_2^2(m_a, m_{abc}; \m_{ab})} .
\end{equation}
According to the spherical law of sines the combinations
\begin{equation} \label{eq:spherical law of sines}
\frac{\sin \theta_{abc}^a(\m_{ab})}{w_{abc}^a} =
\frac{\sin \theta_{ab}^a(m_{abc})}{w_{ab}^a} =
\frac{\sin \theta_{abc}^{ab}(m_a)}{w_{abc}^{ab}} =
\inva_1(\m_{ab}, \m_{bc}),
\end{equation}
corresponds to a dimensionless invariant with the value
\begin{equation} \label{eq:spherical law of sines invariant}
\inva_1(\m_{ab}, \m_{bc}) = \frac{m_a \m_{ab} m_{abc} \volume_3(m_{abc}; m_a, m_b, m_c)}{\volume_2(m_{abc}, \m_{ab}) \volume_2(m_a, \m_{ab}) \volume_2(m_{abc}, m_a)} .
\end{equation}

\subsection{Three-body \PSlong} \label{sec:three-body PS}

\begin{figure}
\begin{panels}{2}
\includepgf*{three-body-limits}
\caption{Invariant masses.} \label{fig:three-body PS masses}
\panel
\includepgf*{three-body-phase-space}
\caption{Angular variables.} \label{fig:three-body PS angles}
\end{panels}
\caption[Diagrams of the three-body \PSlong]{
Diagrams of the three-body \PS parameterised using invariant masses \eqref{eq:three-body PS masses} in panel \subref{fig:three-body PS masses} and angular variables \eqref{eq:three-body PS angles} in panel \subref{fig:three-body PS angles}.
As indicated by the intersecting invariant masses appearing in panel \subref{fig:three-body PS masses} the integration limits in this parameterisation are non-trivial \eqref{eq:three-body integration limits}.
Therefore, we have additionally indicated the quantities that enter the integration limits \eqref{eq:three-body integration limits ingredients} for the integral over $\m_{bc}$.
} \label{fig:three-body PS}
\end{figure}

Using the recursion relation \eqref{eq:recursion relation two-body} the differential three-body \PS can be constructed from a pair of two-body \PSs
\begin{equation} \label{eq:three-body PS}
\d^5 \Phi_3(m_{abc}; m_a, m_b, m_c; m_d, m_e) = \d^2 \Phi_2(m_{abc}; \m_{ab}, m_c; m_d, m_e) \d^3 \Phi_2^\prime(\m_{ab}; m_a, m_b) .
\end{equation}
With the explicit expression of the augmented two-body \PS differential \eqref{eq:augmented two-body PS differential} and the differential two-body \PS \eqref{eq:differential two-body PS} the differential three-body \PS reads \cite{Kinematics:2022pth}
\begin{multline}
\d^5 \Phi_3(m_{abc}; m_a, m_b, m_c; m_d, m_e) = 2 \volume_2(m_{abc}) \frac{\volume_2(\m_{ab})}{\m_{ab}^2} \dfourpi \m_{ab}^2 \\
\dfourpi^2 \Omega_{abcd}^a(\m_{ab}; m_{abc}) \dfourpi^2 \Omega_{abcde}^{ab}(m_{abc}; \m_{abcd}) ,
\end{multline}
The product of the two solid angles differentials \eqref{eq:solid angle differential} can then be expressed as an internal polar opening angle differential \eqref{eq:polar angle differential} and an external Euler angle differential
\begin{equation} \label{eq:Euler angle differential}
\dfourpi^3 \Omega_{abcde;abcd}^{a;ab}(m_{abc}) = \dfourpi \phi_{abcd}^a(\m_{ab}, m_{abc}) \dfourpi^2 \Omega_{abcde}^{ab}(m_{abc}; \m_{abcd}),
\end{equation}
that captures the orientation of the process in three-space and using the integrals \eqref{eq:solid angle differential integrated} integrates to
\begin{equation} \label{eq:Euler angle differential integral}
\iiint \dfourpi^3 \Omega_{abcde;abcd}^{a;ab}(m_{abc}) = \frac{1}{8\pi} .
\end{equation}
The Euler angle differential \eqref{eq:Euler angle differential} is symmetric under the simultaneous exchange of $m_a$ with $\m_{abcde}$ and $\m_{ab}$ with $\m_{abcd}$ as can be seen in the two different parameterisations depicted in \cref{fig:euler}.
Finally, the differential three-body \PS reads
\begin{equation} \label{eq:three-body PS angles}
\d^5 \Phi_3(m_{abc}; m_d, m_e) = 2 \volume_2(m_{abc}) \frac{\volume_2(\m_{ab})}{\m_{ab}^2} \dfourpi \m_{ab}^2 \dfourpi \cos \theta_{abc}^a(\m_{ab}) \dfourpi^3 \Omega_{abcde;abcd}^{a;ab}(m_{abc}) ,
\end{equation}
after integrating over the external Euler angle differential \eqref{eq:Euler angle differential integral} it becomes
\begin{equation} \label{eq:three-body PS angles integrated}
\d^2 \Phi_3(m_{abc}) = \frac{\volume_2(m_{abc})}{4\pi} \frac{\volume_2(\m_{ab})}{\m_{ab}^2} \dfourpi \m_{ab}^2 \dfourpi \cos \theta_{abc}^a(\m_{ab}) ,
\end{equation}
where the two-body integration limits \eqref{eq:two-body limits} are given by
\begin{align} \label{eq:two-body integration limits}
\m_{ab}^{\min} &= m_a + m_b, &
\m_{ab}^{\max} &= m_{abc} - m_c .
\end{align}
The diagram of the three-body \PS in this parameterisation is presented in \cref{fig:three-body PS angles}.
Using the differential of the spherical law of cosines \eqref{eq:three-body derivative}
it is possible to re-write the differential three-body \PS \eqref{eq:three-body PS angles} as
\footnote{
We compensate any sign that might appear in the \PSs by swapping the integration limits.
}
\begin{equation} \label{eq:three-body PS masses}
\d^5 \Phi_3(m_{abc}; m_d, m_e) = \dfourpi \m_{ab}^2 \dfourpi \m_{bc}^2 \dfourpi^3 \Omega_{abcde;abcd}^{a;ab}(m_{abc}) .
\end{equation}
Hence, when parameterising the differential three-body \PS with invariant mass squares it is constant.
After integrating over the external Euler angle differential \eqref{eq:Euler angle differential integral} the differential three-body \PS reads
\begin{equation} \label{eq:three-body PS masses integrated}
\d^2 \Phi_3(m_{abc}) = \frac{1}{8 \pi} \dfourpi \m_{ab}^2 \dfourpi \m_{bc}^2 .
\end{equation}
The three-body integration limits \eqref{eq:three-body integration limits} of the integration over $\m_{bc}$ are given as a function of $\m_{ab}$.
The remaining two-body integration limits are given by \eqref{eq:two-body integration limits}.
The diagram of the three-body \PS in this parameterisation is shown together with the quantities \eqref{eq:three-body integration limits ingredients} that enter the integration limits \eqref{eq:three-body integration limits} in \cref{fig:three-body PS masses}.

\subsection{Three-body processes}

\begin{figure}
\begin{panels}{3}
\includepgf{three-body-distribution}
\caption{Three-body} \label{fig:three-body PS distribution}
\panel
\includepgf{four-body-distribution-1}
\caption{Four-body two-particles} \label{fig:four-body PS distribution 1}
\panel
\includepgf{four-body-distribution-2}
\caption{Four-body three-particles} \label{fig:four-body PS distribution 2}
\end{panels}
\caption[Three- and four-body \PSlong \PDFslong]{
Comparison between the analytical and $\MC$ generated invariant mass distributions of the three-body \PS $\PDFs$ in panel \subref{fig:three-body PS distribution} and the two possible four-body \PS $\PDFs$ in panels \subref{fig:four-body PS distribution 1} and \subref{fig:four-body PS distribution 2}, respectively.
Each of the \PS $\PDFs$ are valid for trivial amplitudes and massless colliding and final state particles.
The three-body \PS $\PDFs$ appears in the three-body decay width \eqref{eq:three-body distribution} and the four-body cross section \eqref{eq:four-body cosine derivative cross section}.
The four-body \PS $\PDFs$ appear in the four-body decay width \eqref{eq:four-body distribution} and the five-body cross section \eqref{eq:five-body differential cross section}.
} \label{fig:three- and four body PS distributions}
\end{figure}

In the following we calculate a decay into three final particles and the two-particle scattering into two final particles using the trivial interaction between four scalar particles with Lagrangian and amplitude
\begin{align}\label{eq:four scalar interaction}
\mathcal L_4 &\supset \frac{\lambda}{4!} \phi_{abc} \phi_a \phi_b \phi_c , &
\mathcal A_4 &= \i \lambda ,
\end{align}
in the limit of massless colliding and final state particles.

\paragraph{Decays into three final particles}

Using either of the parameterisations of the three-body \PS after the integration over the external Euler angle differential \eqref{eq:three-body PS angles integrated,eq:three-body PS masses integrated} together with the trivial amplitude \eqref{eq:four scalar interaction} in order to specify the differential decay width \eqref{eq:decay width} to three-body decays allows to calculate the decay of a particle with mass $m_{abc}$ into three final state particles in the rest frame of the decaying particle as a function of an angle and an invariant mass
\begin{equation}
\d^2 \Gamma_3(m_{abc}) = \frac{\abs{\mathcal A_4}^2}{8 \pi} \frac{\volume_2(m_{abc})}{m_{abc}^3 } \frac{\volume_2(\m_{ab})}{\m_{ab}^2} \dfourpi \m_{ab}^2 \dfourpi \cos \theta_{abc}^a(\m_{ab}) ,
\end{equation}
or alternatively as a function of two invariant masses
\begin{equation}
\d^2 \Gamma_3(m_{abc}) = \frac{\abs{\mathcal A_4}^2}{16 \pi m_{abc}^3} \dfourpi \m_{ab}^2 \dfourpi \m_{bc}^2 ,
\end{equation}
For massless final state particles the differential decay width with respect to $\m_{ab}$ can be expressed as three-body \PS \PDF
\begin{align} \label{eq:three-body distribution}
\frac{1}{2\pi} \frac{\m_{ab}}{\Gamma_3}\dv[\Gamma_3]{\m_{ab}} &= f_3^2\left(\frac{\m_{ab}}{m_{abc}}\right), & f_3^2(x) &= 4 x \left(1 - x^2 \right),
\end{align}
where the total decay width is
\begin{equation}
\Gamma_3(m_{abc}) = \frac{\lambda^2 m_{abc}}{(8\pi)^3} ,
\end{equation}
and we have used that due to the normalisation \eqref{eq:tilded normalisation}
\begin{equation} \label{eq:square tilde derivate}
\frac{\dfourpi \m^2}{\d \m} = \frac{\m}{2\pi} .
\end{equation}
A comparison with \MC data for the three-body \PS \PDF is presented in \cref{fig:three-body PS distribution}.

\paragraph{Two-particle scattering into two final particles}

After inserting the differential three-body \PS in either of the parameterisations \eqref{eq:three-body PS angles} or \eqref{eq:three-body PS masses} into the differential scattering cross section \eqref{eq:cross section} of two particles with masses $m_{abc}$ and $m_c$ into two particles with masses $m_a$ and $m_b$ and integrating over the \COM energy $m_{ab}$ and the external Euler angle differential \eqref{eq:Euler angle differential integral} it reads
\begin{equation}
\d \sigma_3(m_{abc}, m_c; m_{ab}; m_a, m_b) = \frac{\abs{\mathcal A_4}^2}{8 \volume_2(m_{abc})} \frac{\volume_2(m_{ab})}{m_{ab}^2} \dfourpi \cos \theta_{abc}^a(m_{ab}) ,
\end{equation}
or alternatively
\begin{equation}
\d \sigma_3(m_{abc}, m_c; m_{ab}; m_a, m_b) = \frac{\abs{\mathcal A_4}^2}{16\volume_2^2(m_{abc})} \dfourpi \m_{bc}^2 .
\end{equation}
For the production via the trivial four scalar interaction \eqref{eq:four scalar interaction}
and in the limit of vanishing particle masses the total cross section for the \COM energy $m_{ab}$ is then
\begin{equation}
\sigma_3(m_{ab}) = \frac{\lambda^2}{16 \pi m_{ab}^2} .
\end{equation}

\section{Four-body interactions} \label{sec:four-body interaction}

The invariant mass \eqref{eq:invariant mass} of four particles is given by
\begin{align}
m_{abcd}^2 &= \m_{abcd}^2 := \zeta \vec4 p_{abcd}^2, &
\vec4 p_{abcd} &= \vec4 p_a + \vec4 p_b + \vec4 p_c + \vec4 p_d .
\end{align}
Besides four three-body constraints of the form \eqref{eq:three-body invariant mass} for the invariant masses $\m_{abc}$, $\m_{abd}$, $\m_{acd}$, and $\m_{bcd}$,
which can be used to eliminate the non-canonical invariant masses $\m_{ac}$, $\m_{bd}$, $\m_{acd}$, and $\m_{abd}$,
the four-body constraint can be expressed as
\begin{equation} \label{eq:four-body invariant mass}
m_{abcd}^2 = \m_{abc}^2 + \m_{bcd}^2 + \m_{ad}^2 - \m_{bc}^2 - m_a^2 - m_d^2,
\end{equation}
and can be used to eliminate the non-canonical invariant mass $\m_{ad}$.
The four-body \PS is governed by a four-dimensional parallelotope $\parallel_4(m_{abcd}; m_a, m_b, m_c, m_d)$ which is triangulated in twelve different ways by pentatopes $\triangle_4(m_{abcd}; m_a, m_b, m_c, m_d)$.

\subsection{Parallelotopes and pentatopes}

The four-body Gram determinants \eqref{eq:Gram determinant} have mass-dimension six.
The Gram determinant at the vertex $\kappa_{abcd}^a(\m_{ab}, \m_{abc})$ is given by
\begin{equation}
\gram_3^2(m_a, m_{abcd}; \m_{ab}, \m_{abc}) = \zeta \gram_3^2
\begin{pmatrix*}[r]
\vec4 p_a, \vec4 p_{ab}, \vec4 p_{abc} \\
\vec4 p_{abcd}, \vec4 p_{ab}, \vec4 p_{abc}
\end{pmatrix*}
,
\end{equation}
and a Laplace expansion in terms of minors \eqref{eq:Laplace expansion Gram determinant} results in
\begin{multline} \label{eq:four-body Laplace expansion}
\eta \gram_3^2(m_{abcd}, m_a; \m_{ab}, \m_{abc}) = \gram_1^2(m_{abcd}, m_a) \volume_2^2(\m_{ab}, \m_{abc}) \\
- \gram_1^2(m_{abcd}, \m_{ab}) \gram_2^2(m_a, \m_{ab}; \m_{abc}) - \gram_1^2(m_{abcd}, \m_{abc}) \gram_2^2(m_a, \m_{abc}; \m_{ab}) .
\end{multline}
The four-body \CM determinant \eqref{eq:CM determinant} reads
\begin{equation} \label{eq:four-dimensional volume}
\volume_4^2(m_{abcd}; m_a, m_b, m_c, m_d) = \frac1{16}
\begin{vmatrix*}[l]
\header0 & \header1 & \header1 & \header1 & \header1 & \header1 \\
\header1 & \header0 & m_a^2 & \m_{ab}^2 & \m_{abc}^2 & m_{abcd}^2 \\
\header1 & m_a^2 & \header0 & m_b^2 & \m_{bc}^2 & \m_{bcd}^2 \\
\header1 & \m_{ab}^2 & m_b^2 & \header0 & m_c^2 & \m_{cd}^2 \\
\header1 & \m_{abc}^2 & \m_{bc}^2 & m_c^2 & \header0 & m_d^2 \\
\header1 & m_{abcd}^2 & \m_{bcd}^2 & \m_{cd}^2 & m_d^2 & \header0 \\
\end{vmatrix*} .
\end{equation}
The volume of the four-dimensional parallelotope $\parallel_4(m_{abcd}; m_a, m_b, m_c, m_d)$ is then given by $\volume_4(m_{abcd}; m_a, m_b, m_c, m_d)$.
The \DJ determinant identity \eqref{eq:CM determinant DJ identity square symmetric} allows to express the volume compactly as
\begin{equation}
\volume_4^2(m_{abcd}; m_a, m_b, m_c, m_d) =
- \frac{\volume_2^2(\volume_3(m_a, \m_{bc}, m_d), \volume_3(m_b, \m_{ac}, m_d), \volume_3(m_c, \m_{ab}, m_d))}{\volume_2^2(m_d, \m_{abc})} .
\end{equation}
Limiting cases of the volume relevant for \MST are
\begin{equation}
\volume_4^2 =
\frac1{16}\begin{cases}
\!\left. \begin{aligned}
2 \m_{ab}^2 \m_{ac}^2 \m_{bd}^2 \m_{cd}^2 - \m_{ab}^4 \m_{cd}^4 \\
+ 2 \m_{ac}^2 \m_{ad}^2 \m_{bc}^2 \m_{bd}^2 - \m_{ac}^4 \m_{bd}^4 \\
+ 2 \m_{ab}^2 \m_{ad}^2 \m_{bc}^2 \m_{cd}^2 - \m_{ad}^4 \m_{bc}^4
\end{aligned} \right\} & \text{for } m_a = m_b = m_c = m_d = 0, \\
0 & \text{for } m_a + m_b + m_c + m_d = m_{abcd} .
\end{cases}
\end{equation}
Hence the physical regime is determined by positive volumes.
A Laplace expansion in terms of minors \eqref{eq:Laplace expansion CM determinant} results in
\begin{multline}
\eta \volume_4^2(m_{abcd}; m_a, m_b, m_c, m_d) =
 m_a^2 \volume_3^2(m_b, m_c, m_d)
- \gram_1^2(m_a, m_b) \gram_2^2(m_a, m_b; m_c, m_d) \\
- \gram_1^2(m_a, m_c) \gram_2^2(m_a, m_c; m_b, m_d)
- \gram_1^2(m_a, m_d) \gram_2^2(m_a, m_d; m_b, m_c) .
\end{multline}

\subsection{Angle between decay planes} \label{sec:decay plane angle}

In the rest frame of the invariant mass $\m_{ab}$ the cosine of the angle between two planes defined by the four involved momenta can be calculated to be
\begin{equation}
\cos \phi_{abcd}^a(\m_{ab}, \m_{abc}) := \frac{\vec3 p_a(\m_{ab}) \times \vec3 p_{abc}(\m_{ab})}{\abs{\vec3 p_a(\m_{ab}) \times \vec3 p_{abc}(\m_{ab})}} \cdot
\frac{\vec3 p_{abc}(\m_{ab}) \times \vec3 p_{abcd}(\m_{ab})}{\abs{\vec3 p_{abc}(\m_{ab}) \times \vec3 p_{abcd}(\m_{ab})}} ,
\end{equation}
see also \cref{fig:decay planes}.
It can be identified with the angle between the two tetrahedra $\triangle_3(m_a, \m_{ab}, \m_{abc})$ and $\triangle_3(m_{abcd}, \m_{ab}, \m_{abc})$ at the vertex $\vertex_{abcd}^a(\m_{ab}, \m_{abc})$ within the pentatope $\triangle_4(m_{abcd})$.
Specifying the general equation for the cosine of such an angle \eqref{eq:cosine} to the case of pentatopes results in the frame independent expression
\begin{equation} \label{eq:four-body cosine}
\cos \phi_{abcd}^a(\m_{ab}, \m_{abc}) = \frac{\gram_3^2(m_a, m_{abcd}; \m_{ab}, \m_{abc})}{\volume_3(m_a, \m_{ab}, \m_{abc})\volume_3(m_{abcd}, \m_{ab}, \m_{abc})} .
\end{equation}
Note that this expression is not only invariant under exchange of $m_a$ and $m_{abcd}$ but also under exchange of $\m_{ab}$ and $\m_{abc}$.
Using the Laplace expansion \eqref{eq:four-body Laplace expansion} this equation can be
solved for the invariant mass
\begin{equation}
\m_{bcd}^2 = m_{abcd}^2 + m_a^2 - 2 \frac{g_{abcd}^a(\m_{ab},\m_{abc})}{\volume_2^2(\m_{ab}, \m_{abc})} ,
\end{equation}
where
\begin{multline}
g_{abcd}^a(\m_{ab},\m_{abc}) =
\eta \volume_3(m_a, m_b, m_c) \volume_3(\m_{ab}, m_c, m_d) \cos \phi_{abcd}^a(\m_{ab}, \m_{abc}) \\
+ \gram_1^2(m_a, \m_{ab}) \gram_2^2(m_{abcd}, \m_{ab}; \m_{abc})
+ \gram_1^2(m_a, \m_{abc}) \gram_2^2(m_{abcd}, \m_{abc}; \m_{ab})
.
\end{multline}
The volumes of the tetrahedra appearing here can be expressed in terms of polar opening angles using \eqref{eq:three-body sine}.
The derivative of the invariant mass square $\m_{bcd}^2$ with respect to the cosine of the azimuthal decay plane angle reads then
\begin{equation} \label{eq:four-body cosine derivative}
\dv[\m_{bcd}^2]{\cos \phi_{abcd}^a(\m_{ab}, \m_{abc})} = - 2 \eta \frac{\volume_3(m_a, \m_{ab}, \m_{abc})\volume_3(m_{abcd}, \m_{ab}, \m_{abc})}{\volume_2^2(\m_{ab}, \m_{abc})} .
\end{equation}
Specifying the general equation for the sine of the angle between two simplices \eqref{eq:sine} to the case of pentatopes results in
\begin{equation} \label{eq:four-body sine}
\sin\phi_{abcd}^a(\m_{ab}, \m_{abc}) = \frac{\volume_2(\m_{ab}, \m_{abc})\volume_4(m_a, m_b, m_c, m_d)}{\volume_3(m_a, \m_{ab}, \m_{abc})\volume_3(m_{abcd}, \m_{ab}, \m_{abc})} .
\end{equation}
Therefore, using the differential of the invariant mass with respect to the cosine \eqref{eq:four-body cosine derivative} the differential of the invariant mass with respect to the the azimuthal decay plane angle reads
\begin{equation} \label{eq:four-body derivative}
\dv[\m_{bcd}^2]{\phi_{abcd}^a(\m_{ab}, \m_{abc})} = 2 \eta \frac{\volume_4(m_{abcd}; m_a, m_b, m_c, m_d)}{\volume_2(\m_{ab}, \m_{abc})} .
\end{equation}
The tangent
\begin{equation} \label{eq:four-body tangent}
\tan\phi_{abcd}^a(\m_{ab}, \m_{abc}) =
\frac{\volume_2(\m_{ab}, \m_{abc})\volume_4(m_a, m_b, m_c, m_d)}{\gram_3^2(m_a, m_{abcd}; \m_{ab}, \m_{abc})} .
\end{equation}
allows to express the volume without depending on three-dimensional volumes.

\subsection{Invariant angle differentials}

\begin{figure}
\begin{panels}{2}
\includepgf{solid-angle-masses}\includepgf{solid-angle-angles}
\caption{$\dfourpi^2 \Omega_{abcd}^a(\m_{abc}; \m_{ab})$.} \label{fig:solid angle}
\panel
\includepgf{augmented-masses}\includepgf{augmented-angles}
\caption{$\d^3 \Phi_2^\prime(m_{ab}; m_c, m_d)$.} \label{fig:augmented}
\end{panels}
\caption[Solid angle and augmented two-body \PSlong differentials]{
Comparison between the invariant mass and the angular parameterisation of the solid angle differential in panel \subref{fig:solid angle} as well as the augmented two-body \PS differential in panel \subref{fig:augmented}.
For the solid angle differential the invariant mass parameterisation is given in \eqref{eq:solid angle differential invariant} and the angular parameterisation is given in \eqref{eq:solid angle differential}.
For the augmented two-body \PS differential the invariant mass parameterisation is given in \eqref{eq:augmented two-body PS differential invariant} and the angular parameterisation is given in \eqref{eq:augmented two-body PS differential}.
Particle masses are presented as solid blue lines, invariant masses appearing in the differential as dashed red lines, further invariant masses as dotted orange lines, polar angles as dash-dotted green arc, and azimuthal angles as dash-dot-dotted purple arcs.
The relation between the invariant description of the solid angle differential and frame depend depictions is discussed in \cref{fig:solid angle definition}.
} \label{fig:solid angle and augmented two-body PS}
\end{figure}

\begin{figure}
\begin{panels}{2}
\includepgf{euler-masses}
\caption{Invariant mass parameterisation.} \label{fig:euler masses}
\panel
\includepgf{euler-angles}
\caption{Angular parameterisation.} \label{fig:euler angles}
\end{panels}
\caption[Euler angle differentials]{
Comparison between the invariant mass and the angular parameterisation of the Euler angle differential in panels \subref{fig:euler masses} and \subref{fig:euler angles}, respectively.
The Euler angle differential in the invariant mass parameterisation is given in \eqref{eq:Euler angle differential invariant} and in the angular parameterisation in \eqref{eq:Euler angle differential}.
Particle masses are presented as solid blue lines, invariant masses appearing in the differential as dashed red lines, further invariant masses as dotted orange lines, polar angles as dash-dotted green arc, and azimuthal angles as dash-dot-dotted purple arcs.
} \label{fig:euler}
\end{figure}

Using the differential relations \eqref{eq:three-body derivative,eq:four-body derivative} the solid angle differential \eqref{eq:solid angle differential} can be parameterised using invariant quantities \cite{Byckling:1969sx}
\footnote{
Where we have taken into account that the invariant description of the decay plane angle \eqref{eq:four-body sine} is constrained to have half the domain \eqref{eq:angle domain} of the azimuthal angle \eqref{eq:solid angle domain}.
}
\begin{equation} \label{eq:solid angle differential invariant}
\dfourpi^2 \Omega_{abcd}^a(\m_{abc}; \m_{ab}) = \frac{\m_{ab}^2}{2 \volume_2(\m_{ab})} \frac{\dfourpi \m_{bc}^2 \dfourpi \m_{bcd}^2}{\volume_4(m_{abcd})} .
\end{equation}
Additionally, using the differential two-body \PS \eqref{eq:differential two-body PS} together with this expressions allows to parameterise the augmented two-body \PS differential \eqref{eq:augmented two-body PS differential} in terms of invariants
\begin{equation} \label{eq:augmented two-body PS differential invariant}
\d^3 \Phi_2^\prime(m_{ab}; m_c, m_d) = \frac{\dfourpi \m_{ab}^2 \dfourpi \m_{bc}^2 \dfourpi \m_{bcd}^2}{\volume_4(m_{abcd})} .
\end{equation}
The comparison of the two parameterisations of these differentials in \cref{fig:solid angle and augmented two-body PS} as well as the appearance of the four-dimensional volume illustrates that these differentials are connected to the four-body \PS.
Furthermore, the Euler angle differential \eqref{eq:Euler angle differential} can be invariantly parameterised using the differential relation \eqref{eq:four-body derivative} together with the invariant parameterisation of the solid angle differential \eqref{eq:solid angle differential invariant}
\begin{equation} \label{eq:Euler angle differential invariant}
\dfourpi^3 \Omega_{abcde;abcd}^{a;ab}(\m_{abc}) =
\frac{\m_{abc}^2}{2 \volume_4(\m_{abcd})}
\frac{\dfourpi \m_{bcd}^2 \dfourpi \m_{cd}^2 \dfourpi \m_{cde}^2}{\volume_4(\m_{bcde})} .
\end{equation}
The comparison of the two parameterisations in \cref{fig:euler} show that this differential is connected to the five-body \PS discussed in \cref{sec:five-body PS}.

\subsection{Four-body \PSlong}

\begin{figure}
\begin{panels}{2}
\includepgf*{four-body-phase-space-masses}
\caption{Invariant masses.} \label{fig:four-body PS masses}
\panel
\includepgf*{four-body-phase-space-angles}
\caption{Angular variables.} \label{fig:four-body PS angles}
\end{panels}
\caption[Diagrams of the four-body \PSlong]{
Diagrams of the four-body \PS parameterised using invariant masses \eqref{eq:four-body PS masses} in panel \subref{fig:four-body PS masses} and angular variables \eqref{eq:four-body PS angles} in panel \subref{fig:four-body PS angles}.
} \label{fig:four-body PS}
\end{figure}

Using the recursion relation \eqref{eq:recursion relation}, the differential four-body \PS can be expressed as a function of a differential three-body \PS \eqref{eq:three-body PS masses} together with an augmented two-body \PS differential \eqref{eq:augmented two-body PS differential}
\begin{equation}
\d^8 \Phi_4(m_{abcd}; m_e, m_f) = \d^5 \Phi_3(m_{abcd}; \m_{ab}, m_c, m_d; m_e, m_f) \d^3 \Phi_2^\prime(\m_{ab}; m_c, m_d) ,
\end{equation}
using the invariant parameterisation for the differential three-body \PS \eqref{eq:three-body PS masses} and the augmented two-body \PS differential \eqref{eq:augmented two-body PS differential invariant} the differential four-body \PS reads
\begin{equation} \label{eq:four-body PS masses}
\d^8 \Phi_4(m_{abcd}; m_e, m_f) = \dfourpi \m_{abc}^2 \dfourpi \m_{cd}^2 \frac{\dfourpi \m_{ab}^2 \dfourpi \m_{bc}^2 \dfourpi \m_{bcd}^2}{\volume_4(m_{abcd})} \dfourpi^3 \Omega_{abcdef;abcde}^{ab;abc}(m_{abcd}) ,
\end{equation}
Therefore, the four-body \PS parameterised by invariant masses is inversely proportional to the volume of a four-dimensional parallelotope.
This parameterisation is depicted in \cref{fig:four-body PS masses}.
The explicit appearance of the volume and the non-trivial integration limits of the form \eqref{eq:three-body integration limits} can become problematic for analytic calculations \cite{Kumar:1969jjy,Kumar:1970wm}.
Alternatively, the differential four-body \PS can be constructed from three two-body \PSs
\begin{multline}
\d^8 \Phi_4(m_{abcd}; m_e, m_f) = \\
\d^2 \Phi_2(m_{abcd}; \m_{abc}, m_d; m_e, m_f) \d^3 \Phi_2^\prime(\m_{abc}; \m_{ab}, m_c; m_d, m_e) \d^3 \Phi_2^\prime(\m_{ab}; m_c, m_d) ,
\end{multline}
Using the explicit expressions for the augmented two-body \PS differential \eqref{eq:augmented two-body PS differential} and the differential two-body \PS \eqref{eq:differential two-body PS} the four-body \PS reads
\begin{multline} \label{eq:four-body PS angles}
\d^8 \Phi_4(m_{abcd}; m_e, m_f) = 4 \volume_2(m_{abcd})
\frac{\volume_2(\m_{abc})}{\m_{abc}^2}
\frac{\volume_2(\m_{ab})}{\m_{ab}^2}
\dfourpi \m_{abc}^2
\dfourpi \m_{ab}^2 \\
\dfourpi \cos \theta_{abcd}^{ab}(\m_{abc})
\dfourpi \phi_{abcd}^a(\m_{abc}, \m_{ab})
\dfourpi \cos \theta_{abc}^a(\m_{ab})
\dfourpi^3 \Omega_{abcdef;abcde}^{ab;abc}(m_{abcd}) ,
\end{multline}
This parameterisation is depicted in \cref{fig:four-body PS angles}.
However, further parameterisations have also been studied \cite{Yu:2021dtp}.

\subsection{Four-body processes}

In the following we use a trivial interaction between five scalar particles with Lagrangian and amplitude
\begin{align} \label{eq:four-body Lagrangian}
\mathcal L_5 &\supset \frac{1}{5!\Lambda} \phi_{abcd} \phi_a \phi_b \phi_c \phi_d, &
\mathcal A_5 &= \frac{\i}{\Lambda} ,
\end{align}
in the limit of massless colliding and final state particles.

\paragraph{Decays into three final particles}

The partial decay width \eqref{eq:decay width} into four final state particles via the trivial interaction \eqref{eq:four-body Lagrangian} in the rest frame of the decaying particle using the \PS parameterisation \eqref{eq:four-body PS angles} reads after integration over the external Euler angle differential
\begin{multline}
\d^5 \Gamma_4(m_{abcd}) =
\frac{\abs{\mathcal A_5}^2}{4\pi}
\frac{\volume_2(m_{abcd})}{m_{abcd}^3}
\frac{\volume_2(\m_{abc})}{\m_{abc}^2}
\frac{\volume_2(\m_{ab})}{\m_{ab}^2}
\dfourpi \m_{abc}^2
\dfourpi \m_{ab}^2 \\
\dfourpi \cos \theta_{abcd}^{ab}(\m_{abc})
\dfourpi \phi_{abcd}^a(\m_{abc}, \m_{ab})
\dfourpi \cos \theta_{abc}^a(\m_{ab})
.
\end{multline}
After evaluating four integrals the one-dimensional differential decay widths with respect to the invariant masses $\m_{ab}$ and $\m_{abc}$ read
\begin{align}
\frac{1}{2\pi} \frac{\m_{ab}}{\Gamma_4} \dv[\Gamma_4]{\m_{ab}} &= f_4^2\left(\frac{\m_{ab}}{m_{abcd}}\right), &
\frac{1}{2\pi} \frac{\m_{abc}}{\Gamma_4} \dv[\Gamma_4]{\m_{abc}} &= f_4^3\left(\frac{\m_{abc}}{m_{abcd}}\right) .
\end{align}
where we have used the derivative \eqref{eq:square tilde derivate}.
The four-body \PS \PDF for massless final state particles appearing here are
\begin{align} \label{eq:four-body distribution}
f_4^2(x) &= 12 x \left(1 - x^4 + 2 x^2 \ln x^2 \right), &
f_4^3(x) &= 12 x^3 \left(1 - x^2 \right) ,
\end{align}
and the total decay width is
\begin{equation}
\Gamma_4(m_{abcd}) = \frac{2 m_{abcd}^3}{3 (8\pi)^5 \Lambda^2} .
\end{equation}
A comparison with \MC data is shown in \cref{fig:four-body PS distribution 1,fig:four-body PS distribution 2}, respectively.

\paragraph{Two particle scattering into three final particles}

The differential production cross section \eqref{eq:cross section} of the four-body process \eqref{eq:four-body Lagrangian} in the parameterisation \eqref{eq:four-body PS angles} reads after integration over the \COM energy and the external Euler angle differential \eqref{eq:Euler angle differential integral}
\begin{multline}
\d^7 \sigma_4(m_{abcd}, m_d; m_{abc}) =
\frac{\abs{\mathcal A_5}^2}{4 \volume_2(m_{abcd})}
\frac{\volume_2(m_{abc})}{m_{abc}^2}
\frac{\volume_2(\m_{ab})}{\m_{ab}^2}
\dfourpi \m_{ab}^2 \\
\dfourpi \cos \theta_{abcd}^{ab}(m_{abc})
\dfourpi \phi_{abcd}^a(m_{abc}, \m_{ab})
\dfourpi \cos \theta_{abc}^a(\m_{ab})
,
\end{multline}
Hence in the limit of vanishing particle masses the distribution of the differential production cross section is given by the three-body \PS \PDF \eqref{eq:three-body distribution}
\begin{equation} \label{eq:four-body cosine derivative cross section}
\frac{1}{2\pi} \frac{\m_{ab}}{\sigma_4}\dv[\sigma_4]{\m_{ab}} = f_3^2\left(\frac{\m_{ab}}{m_{abc}}\right),
\end{equation}
where the total production cross section
\begin{equation}
\sigma_4 = \frac{1}{(8\pi)^3 \Lambda^2} ,
\end{equation}
is independent of the \COM energy.
A comparison with the \MC generated data is presented in \cref{fig:three-body PS distribution}.

\section{Five-body interactions} \label{sec:five-body interaction}

The invariant mass \eqref{eq:invariant mass} of five particles is given by
\begin{align}
m_{abcde}^2 &= \m_{abcde}^2 := \zeta \vec4 p_{abcde}^2, &
\vec4 p_{abcde} &= \vec4 p_a + \vec4 p_b + \vec4 p_c + \vec4 p_d + \vec4 p_e .
\end{align}
Besides the three-body constraints of the form \eqref{eq:three-body invariant mass} and the four-body constraints of the form \eqref{eq:four-body invariant mass} the five-body constraint can be expressed as
\begin{equation} \label{eq:five-body invariant mass}
m_{abcde}^2 = \m_{abcd}^2 + \m_{bcde}^2 + \m_{ae}^2 - \m_{bcd}^2 - m_a^2 - m_e^2,
\end{equation}
and can be used to eliminate the non-canonical invariant mass $\m_{ae}$.

\subsection{Parallelotopes and five-simplex}

The five-body Gram determinants \eqref{eq:Gram determinant} have mass-dimension eight.
The Gram determinant at the vertex $\kappa_{abcde}^a(\m_{ab}, \m_{abc}, \m_{abcd})$ is given by
\begin{equation} \label{eq:five-body Gram determinant}
\gram_4^2(m_a, m_{abcde}; \m_{ab}, \m_{abc}, \m_{abcd}) =
\gram_4^2
\begin{pmatrix*}[r]
\vec4 p_a, \vec4 p_{ab}, \vec4 p_{abc}, \vec4 p_{abcd} \\
\vec4 p_{abcde}, \vec4 p_{ab}, \vec4 p_{abc}, \vec4 p_{abcd}
\end{pmatrix*}
,
\end{equation}
and a Laplace expansion in terms of minors \eqref{eq:Laplace expansion Gram determinant} results in
\begin{equation} \label{eq:five-body Laplace expansion}
\begin{split}
\eta \gram_4^2(m_{abcde}, m_a; \m_{ab}, \m_{abc}, \m_{abcd}) ={}&
\gram_1^2(m_{abcde}, m_a) \volume_3^2(\m_{ab}, \m_{abc}, \m_{abcd}) \\
&- \gram_1^2(m_{abcde}, \m_{ab}) \gram_3^2(m_a, \m_{ab}; \m_{abc}, \m_{abcd}) \\
&- \gram_1^2(m_{abcde}, \m_{abc}) \gram_3^2(m_a, \m_{abc}; \m_{ab}, \m_{abcd}) \\
&- \gram_1^2(m_{abcde}, \m_{abcd}) \gram_3^2(m_a, \m_{abcd}; \m_{ab}, \m_{abc}) .
\end{split}
\end{equation}
The five-body \CM determinant \eqref{eq:CM determinant} is
\begin{equation} \label{eq:five-body volume}
\volume_5^2(m_{abcde}; m_a, m_b, m_c, m_d, m_e) = \frac1{32}\begin{vmatrix*}[l]
\header0 & \header1 & \header1 & \header1 & \header1 & \header1 & \header1 \\
\header1 & \header0 & m_a^2 & \m_{ab}^2 & \m_{abc}^2 & \m_{abcd}^2 & m_{abcde}^2 \\
\header1 & m_a^2 & \header0 & m_b^2 & \m_{bc}^2 & \m_{bcd}^2 & \m_{bcde}^2 \\
\header1 & \m_{ab}^2 & m_b^2 & \header0 & m_c^2 & \m_{cd}^2 & \m_{cde}^2 \\
\header1 & \m_{abc}^2 & \m_{bc}^2 & m_c^2 & \header0 & m_d^2 & \m_{de}^2 \\
\header1 & \m_{abcd}^2 & \m_{bcd}^2 & \m_{cd}^2 & m_d^2 & \header0 & m_e^2 \\
\header1 & m_{abcde}^2 & \m_{bcde}^2 & \m_{cde}^2 & \m_{de}^2 & m_e^2 & \header0 \\
\end{vmatrix*} .
\end{equation}
Therefore, the five-dimensional parallelotope is parameterised by nine canonical invariant masses.

\subsection{Angle and constraint}

Specifying the general equation for the cosine of an angle \eqref{eq:cosine} to five dimension results in
\begin{equation}
\cos \psi_{abcde}^a(\m_{ab}, \m_{abc}, \m_{abcd}) = \frac{\gram_4^2(m_a, m_{abcd}; \m_{ab}, \m_{abc}, \m_{abcd})}{\volume_4(m_a, \m_{ab}, \m_{abc}, \m_{abcd}) \volume_4(m_{abcde}, \m_{ab}, \m_{abc}, \m_{abcd})} ,
\end{equation}
and similarly, the sine \eqref{eq:sine} reads
\begin{equation}
\sin \psi_{abcde}^a(\m_{ab}, \m_{abc}, \m_{abcd}) = \frac{\volume_3(\m_{ab}, \m_{abc}, \m_{abcd}) \volume_5(m_{abcde})}{\volume_4(m_a, \m_{ab}, \m_{abc}, \m_{abcd}) \volume_4(m_{abcde}, \m_{ab}, \m_{abc}, \m_{abcd})} .
\end{equation}
Since the five-dimensional volume \eqref{eq:five-body volume} must vanish in a four-dimensional \MST \cite{Byers:1964ryc} only eight of the nine canonical invariant masses are independent, see also the discussion in \cref{sec:invariants}.
Therefore, we require that
\begin{equation}
\gram_4^2(m_a, m_{abcde}; \m_{ab}, \m_{abc}, \m_{abcd}) = \volume_4(m_a, \m_{ab}, \m_{abc}, \m_{abcd}) \volume_4(m_{abcde}, \m_{ab}, \m_{abc}, \m_{abcd}) .
\end{equation}
Using the Laplace expansion \eqref{eq:five-body Laplace expansion} to solve this expression for the invariant mass $\m_{bcde}$ results in the constraint
\begin{equation} \label{eq:five-body constraint}
\m^2_{bcde} = m_a^2 + m^2_{abcde} - 2 \frac{g_{abcde}^a(\m_{ab}, \m_{abc}, \m_{abcd})}{\volume_3^2(\m_{ab},\m_{abc},\m_{abcd})} ,
\end{equation}
where
\begin{equation}
\begin{split}
g_{abcde}^a(\m_{ab}, \m_{abc}, \m_{abcd}) ={}&
\eta \volume_4(m_a, \m_{ab}, \m_{abc}, \m_{abcd}) \volume_4(m_{abcde}, \m_{ab}, \m_{abc}, \m_{abcd})\\
&+ \gram_1^2(m_{abcde}, \m_{ab}) \gram_3^2(m_a, \m_{ab}; \m_{abc}, \m_{abcd}) \\
&+ \gram_1^2(m_{abcde}, \m_{abc}) \gram_3^2(m_a, \m_{abc}; \m_{ab}, \m_{abcd}) \\
&+ \gram_1^2(m_{abcde}, \m_{abcd}) \gram_3^2(m_a, \m_{abcd}; \m_{ab}, \m_{abc})
.
\end{split}
\end{equation}
The volumes $\volume_3$ and $\volume_4$ can be expressed in terms of polar opening angles \eqref{eq:three-body sine} and azimuthal decay plane angles \eqref{eq:four-body tangent}, respectively.
This constraint allows to eliminate the invariant mass~$\m_{bcde}$.

\subsection{Five-body \PSlong} \label{sec:five-body PS}

\begin{figure}
\begin{panels}{.45}
\includepgf{five-body-phase-space-masses}
\caption{Invariant masses} \label{fig:five-body PS masses}
\panel{.55}
\includepgf{five-body-phase-space-angles}
\caption{Angular variables} \label{fig:five-body PS angles}
\end{panels}
\caption[Diagrams of the five-body \PSlong]{
Diagrams of the five-body \PS parameterised using invariant masses \eqref{eq:five-body PS masses} in panel \subref{fig:five-body PS masses} and angular variables \eqref{eq:five-body PS angles} in panel \subref{fig:five-body PS angles}.
While the parallelotope is parameterised by nine variables the \PS in \MST is constrained since the five-dimensional volume \eqref{eq:five-body volume} must vanish.
Therefore one of the invariant masses (dotted orange line) can be eliminated using the constraint \eqref{eq:five-body constraint}.
} \label{fig:five-body PS}
\end{figure}

The differential five-body \PS can be constructed from a differential three-body \PS \eqref{eq:three-body PS} together with two augmented two-body \PS differentials \eqref{eq:augmented two-body PS differential} using the recursion relation \eqref{eq:recursion relation}
\begin{multline}
\d^{11} \Phi_5(m_{abcde}; m_f, m_g) = \d^5 \Phi_3(\m_{abcde}; m_{abc}, m_d, m_e; m_f, m_g) \\
\d^3 \Phi_2^\prime(\m_{abc}; \m_{ab}, m_c; m_d, m_e) \d^3 \Phi_2^\prime(\m_{ab}; m_c, m_d) .
\end{multline}
After inserting the invariant parameterisations of the differential three-body \PS \eqref{eq:three-body PS masses} and the invariant expression for the augmented two-body \PS differential \eqref{eq:augmented two-body PS differential invariant} and integrating over the external Euler angle differential \eqref{eq:Euler angle differential integral} the differential five-body \PS
\begin{equation} \label{eq:five-body PS masses}
\d^8 \Phi_5(m_{abcde}) = \frac{1}{8\pi} \dfourpi \m_{abcd}^2 \dfourpi \m_{de}^2
\frac{\dfourpi \m_{abc}^2 \dfourpi \m_{cd}^2 \dfourpi \m_{cde}^2 }{\volume_4(\m_{bcde})}
\frac{\dfourpi \m_{ab}^2 \dfourpi \m_{bc}^2 \dfourpi \m_{bcd}^2}{\volume_4(\m_{abcd})} ,
\end{equation}
depends only on invariant masses.
The diagram of the five-body \PS in this parameterisation is shown in \cref{fig:five-body PS masses}.
The appearance of four-dimensional volumes \eqref{eq:four-dimensional volume} and non-trivial integration limits \eqref{eq:three-body integration limits} renders this parameterisation impractical for analytical calculations.
Alternatively, the differential five-body \PS can be constructed from four differential two-body \PSs
\begin{multline}
\d^{11} \Phi_5(m_{abcde}; m_f, m_g) =
\d^2 \Phi_2(m_{abcde}; \m_{abcd}, m_e; m_f, m_g)
\d^3 \Phi_2^\prime(\m_{abcd}; \m_{abc}, m_d; m_e, m_f) \\
\d^3 \Phi_2^\prime(\m_{abc}; \m_{ab}, m_c; m_d, m_e)
\d^3 \Phi_2^\prime(\m_{ab}; m_c, m_d) .
\end{multline}
After inserting the explicit expressions for the augmented two-body \PS differential \eqref{eq:augmented two-body PS differential} and the differential two-body \PS \eqref{eq:differential two-body PS} and integrating over the external Euler angle differential \eqref{eq:Euler angle differential integral} the differential five-body \PS reads
\begin{multline} \label{eq:five-body PS angles}
\d^8 \Phi_5(m_{abcde}) =
\frac{\volume_2(m_{abcde})}{\pi}
\frac{\volume_2(\m_{abcd})}{\m_{abcd}^2}
\frac{\volume_2(\m_{abc})}{\m_{abc}^2}
\frac{\volume_2(\m_{ab})}{\m_{ab}^2}
\dfourpi \m_{abcd}^2
\dfourpi \m_{abc}^2
\dfourpi \m_{ab}^2 \\
\dfourpi \cos \theta_{abcde}^{abc}(\m_{abcd})
\dfourpi \phi_{abcde}^{ab}(\m_{abcd}, \m_{abc})
\dfourpi \cos \theta_{abcd}^{ab}(\m_{abc}) \\
\dfourpi \phi_{abcd}^a(\m_{abc}, \m_{ab})
\dfourpi \cos \theta_{abc}^a(\m_{ab})
.
\end{multline}
The \PS diagram of this parameterisation is shown in \cref{fig:five-body PS angles}.

\subsection{Five-body processes}

In the following we work with the trivial interaction of six scalar particles with Lagrangian and amplitude
\begin{align} \label{eq:five-body Lagrangian}
\mathcal L_6 &\supset \frac{1}{6!\Lambda^2} \phi_{abcde} \phi_a \phi_b \phi_c \phi_d \phi_e , &
\mathcal A_6 &= \frac{\i}{\Lambda^2} .
\end{align}
and massless colliding and final state particles.

\paragraph{Decays into five final particles}

\begin{figure}
\begin{panels}{3}
\includepgf{five-body-distribution-1}
\caption{Two particles} \label{fig:five-body PS distribution 1}
\panel
\includepgf{five-body-distribution-2}
\caption{Three particles} \label{fig:five-body PS distribution 2}
\panel
\includepgf{five-body-distribution-3}
\caption{Four particles} \label{fig:five-body PS distribution 3}
\end{panels}
\caption[Five-body \PSlong \PDFslong]{
Comparison between the analytical and \MC generated five-body \PS \PDFs for massless final state particles \eqref{eq:five-body distribution} and for invariant masses of two, three, and four particle in panels \subref{fig:five-body PS distribution 1}, \subref{fig:five-body PS distribution 2}, and \subref{fig:five-body PS distribution 3}, respectively.
These \PDFs appears in the five-body decay width \eqref{eq:five-body decy width distribution} and the six-body cross section.
} \label{fig:five-body PS distributions}
\end{figure}

The differential decay-width \eqref{eq:decay width} using the parameterisation \eqref{eq:five-body PS angles} of the differential five-body \PS reads after integrating over the external Euler angle differential \eqref{eq:Euler angle differential integral}
\begin{multline}
\d^8 \Gamma_5(m_{abcde}) =
\frac{\abs{\mathcal A_6}^2}{2\pi}
\frac{\volume_2(m_{abcde})}{m_{abcde}^3}
\frac{\volume_2(\m_{abcd})}{\m_{abcd}^2}
\frac{\volume_2(\m_{abc})}{\m_{abc}^2}
\frac{\volume_2(\m_{ab})}{\m_{ab}^2}
\dfourpi \m_{abcd}^2
\dfourpi \m_{abc}^2
\dfourpi \m_{ab}^2 \\
\dfourpi \cos \theta_{abcde}^{abc}(\m_{abcd})
\dfourpi \phi_{abcde}^{ab}(\m_{abcd}, \m_{abc})
\dfourpi \cos \theta_{abcd}^{ab}(\m_{abc}) \\
\dfourpi \phi_{abcd}^a(\m_{abc}, \m_{ab})
\dfourpi \cos \theta_{abc}^a(\m_{ab})
.
\end{multline}
The distribution of the one-dimensional differential decay width with respect to the invariant masses $\m_{ab}$, $\m_{abc}$, and $\m_{abcd}$ are
\begin{subequations}
\label{eq:five-body decy width distribution}
\begin{align}
\frac{1}{2\pi} \frac{\m_{ab}}{\Gamma_5} \dv[\Gamma_5]{\m_{ab}} &= f_5^2\left(\frac{\m_{ab}}{m_{abcde}}\right), &
\frac{1}{2\pi} \frac{\m_{abc}}{\Gamma_5} \dv[\Gamma_5]{\m_{abc}} &= f_5^3\left(\frac{\m_{abc}}{m_{abcde}}\right), \\
\frac{1}{2\pi} \frac{\m_{abcd}}{\Gamma_5} \dv[\Gamma_5]{\m_{abcd}} &= f_5^4\left(\frac{\m_{abcd}}{m_{abcde}}\right) .
\end{align}
\end{subequations}
where the five-body \PS \PDFs for massless final state particles are
\begin{subequations} \label{eq:five-body distribution}
\begin{align}
f_5^2(x) &= 24 x^5 \left(1 - x^2\right), &
f_5^3(x) &= 72 x^3 \left(1 - x^4 + 2 x^2 \ln x^2\right), \\
f_5^4(x) &= 24 x \mathrlap{\left[1 + 9 (x^2 - x^4) - x^6 + 6 (x^2 + x^4) \ln x^2\right] ,}
\end{align}
\end{subequations}
and the total decay-width is
\begin{equation}
\Gamma_5(m_{abcde}) = \frac{2 m_{abcde}^5}{3^2 (8\pi)^7 \Lambda^4} .
\end{equation}
A comparison with \MC data is presented in \cref{fig:five-body PS distributions}.

\paragraph{Two-particle scattering into four final particles}

After inserting the parameterisation \eqref{eq:five-body PS angles} of the differential five-body \PS into the differential production cross section \eqref{eq:cross section} and integrating over the external Euler angle differential \eqref{eq:Euler angle differential integral} and the \COM energy $m_{abcd}$ it reads
\begin{multline}
\d^7 \sigma_5(m_{abcde}, m_e; m_{abcd}) =
\frac{\abs{\mathcal A_6}^2}{2\volume_2(m_{abcde})}
\frac{\volume_2(m_{abcd})}{m_{abcd}^2}
\frac{\volume_2(\m_{abc})}{\m_{abc}^2}
\frac{\volume_2(\m_{ab})}{\m_{ab}^2}
\dfourpi \m_{abc}^2
\dfourpi \m_{ab}^2 \\
\dfourpi \cos \theta_{abcde}^{abc}(m_{abcd})
\dfourpi \phi_{abcde}^{ab}(\m_{abcd}, m_{abc})
\dfourpi \cos \theta_{abcd}^{ab}(\m_{abc}) \\
\dfourpi \phi_{abcd}^a(\m_{abc}, \m_{ab})
\dfourpi \cos \theta_{abc}^a(\m_{ab})
,
\end{multline}
The total production cross section for the trivial amplitude \eqref{eq:five-body Lagrangian} and massless initial and final state particles is then
\begin{equation}
\sigma_5(m_{abcde}) = \frac{2 m_{abcd}^2}{3 (8\pi)^5 \Lambda^4}
\end{equation}
The two- and three-particle distributions of the one-dimensional differential cross section are
\begin{align} \label{eq:five-body differential cross section}
\frac{1}{2\pi} \frac{\m_{ab}}{\sigma_5} \dv[\sigma_5]{\m_{ab}} &= f_4^2\left(\frac{\m_{ab}}{m_{abcd}}\right) , &
\frac{1}{2\pi} \frac{\m_{abc}}{\sigma_5} \dv[\sigma_5]{\m_{abc}} &= f_4^3\left(\frac{\m_{abc}}{m_{abcd}}\right) .
\end{align}
where the four-body \PS \PDFs are given in \eqref{eq:four-body distribution}.
A comparison with \MC data is shown in \cref{fig:four-body PS distribution 1,fig:four-body PS distribution 2}.

\section{Multi-body interactions}\label{sec:multi-body interaction}

The volume of the parallelotope that governs an $n$-body interactions \eqref{eq:CM determinant} can be parameterised by $\flatfrac{(n - 2)(n + 1)}2$ canonical invariant masses.
Since volumes with dimensions larger than four have to vanish \cite{Byers:1964ryc}, $\flatfrac{(n - 4) (n - 3)}2$ constraints of the form \eqref{eq:five-body constraint} have to be considered for $n > 2$.
Therefore the differential $n$-dimensional \PS is parameterised by $3 n - 7$ independent internal \DOFs.

\subsection{Phase space}

\begin{figure}
\begin{panels}{.4}
\includepgf{n-body-phase-space-masses}
\caption{Invariant masses.} \label{fig:n-body PS masses}
\panel{.6}
\includepgf{n-body-phase-space-angles}
\caption{Angular variables.} \label{fig:n-body PS angles}
\end{panels}
\caption[Diagrams of the $n$-body \PSlong]{
Diagrams of the $n$-body \PS when parameterised using invariant masses \eqref{eq:n-body PS masses} in panel \subref{fig:n-body PS masses} and angular variables \eqref{eq:n-body PS angles} in panel \subref{fig:n-body PS angles}.
The on-shell masses of the final state particles are shown as solid blue lines, the integrals over invariant masses as dashed red lines, further invariant masses as dotted orange lines, polar opening angles as dashed-dotted green arcs, and azimuthal decay-plane angles as dashed-double-dotted purple arcs.
Both depictions show that $n-3$ different four-body \PSs,
spanned by variables labeled with an $i$,
are involved,
while the last two external particles span a three-body \PS
that contains variables labeled with an $n$.
See \cref{fig:two-body PS,fig:three-body PS} for a depiction of these simple \PSs.
}
\end{figure}

Using the recursion relation \eqref{eq:recursion relation} repeatedly the differential $n$-body \PS can be written in terms of a differential three-body \PS \eqref{eq:three-body PS} and a product of augmented two-body \PS differentials \eqref{eq:augmented two-body PS differential}
\begin{multline} \label{eq:PS decompositions}
\d^{3n-4} \Phi_n(m_{a\cdots n}; m_{n+1}, \dots) =
\d^5 \Phi_3(m_{a\cdots n}; \m_{a\cdots n-2}, m_{n-1}, m_n; m_{n+1}, \dots) \\
\prod_{i=b}^{n-2} \d^3 \Phi_2^\prime(\m_{a\cdots i};m_{i+1}, m_{i+2}).
\end{multline}
After using the invariant expressions for the differential three-body \PS \eqref{eq:three-body PS masses} as well as for the augmented two-body \PS differentials \eqref{eq:augmented two-body PS differential invariant} and integrating over the external Euler angle differential \eqref{eq:Euler angle differential integral} the invariant expression for the differential $n$-body \PS reads
\begin{equation} \label{eq:n-body PS masses}
\d^{3n-7} \Phi_n(m_{a\cdots n})
= \frac{1}{8\pi} \dfourpi \m_{a\cdots n-1}^2 \dfourpi \m_{n,n-1}^2
\prod_{i=b}^{n-2} \frac{
\dfourpi \m_{a\cdots i}^2
\dfourpi \m_{i,i+1}^2
\dfourpi \m_{i,i+1,i+2}^2
}{\volume_4(\m_{a\cdots i-1}, m_i, m_{i+1}, m_{i+2})},
\end{equation}
The diagram of the differential $n$-body \PS in this parameterisation is presented in \cref{fig:n-body PS masses}.
In particular this expression recovers the frame independent parameterisation for the differential three-body \PS \eqref{eq:three-body PS masses}, the differential four-body \PS \eqref{eq:four-body PS masses}, and the differential five-body \PS \eqref{eq:five-body PS masses}.
While this expression depends only on integrals over squares of invariant masses, the appearance of four-dimensional volumes \eqref{eq:four-dimensional volume} and non-trivial integration limits of the type \eqref{eq:three-body integration limits} render the expression impractical for analytical calculations.
Alternatively, the differential $n$-body \PS \eqref{eq:PS decompositions} can be be rewritten in terms of differential two-body \PSs
\begin{multline}
\d^{3n-4} \Phi_n(m_{a\cdots n}; m_{n+1}, \dots) =
\d^2 \Phi_2(m_{a\cdots n}; \m_{a\cdots n-1}, m_n; m_{n+1}, \dots) \\
\prod_{i=b}^{n-1} \d^3 \Phi_2^\prime(\m_{a\cdots i}; m_{i+1}, m_{i+2}) .
\end{multline}
Using the explicit expressions for the augmented two-body \PS differential \eqref{eq:augmented two-body PS differential} and the differential two-body \PS \eqref{eq:differential two-body PS}
and integrating over the external Euler angle differential \eqref{eq:Euler angle differential integral} the differential $n$-body \PS reads
\begin{multline} \label{eq:n-body PS angles}
\d^{3n-7} \Phi_n(m_{a\cdots n}) =
\frac{\volume_2(m_{a\cdots n})}{4\pi}
\frac{\volume_2(\m_{a\cdots n-1})}{m_{a\cdots n-1}^2} \dfourpi \m_{a\cdots n-1}^2
\dfourpi \cos \theta_{a\cdots n-2}^{a\cdots n}(\m_{a\cdots n-1}) \\
\prod_{i=b}^{n-2}
2 \frac{\volume_2(\m_{a\cdots i})}{\m_{a\cdots i}^2}
\dfourpi \m_{a\cdots i}^2
\dfourpi \phi_{a\cdots i-1}^{a\cdots i+2}(\m_{a\cdots i}, \m_{a\cdots i+1})
\dfourpi \cos \theta_{a\cdots i-1}^{a\cdots i+1}(\m_{a\cdots i})
,
\end{multline}
where the areas \eqref{eq:area} are given in the notation \eqref{eq:multi-body notation} and the differentials are normed according to \eqref{eq:tilded normalisation}.
In particular this expression recovers the differential three-body \PS \eqref{eq:three-body PS angles}, the differential four-body \PS \eqref{eq:four-body PS angles}, and the differential five-body \PS \eqref{eq:five-body PS angles} in the parameterisation depending on angular variables.
The diagram of the $n$-body \PS in this parameterisation is depicted in \cref{fig:n-body PS angles}.
Therefore, the differential $n$-body \PS can be expressed as a product of $n-1$ areas \eqref{eq:area} together with  $n-2$ invariant mass differentials, $n-2$ differential of polar opening angles \eqref{eq:three-body cosine}, and $n-3$ differentials of azimuthal decay plane angles \eqref{eq:four-body cosine}.
The optimal set of integration variables can be identified by comparing the diagram in \cref{fig:n-body PS angles} with the Feynman diagram of the corresponding amplitude.

\section{Conclusion} \label{sec:conculsion}

\begin{figure}
\begin{panels}{.16}
\includepgf*{two-body-feynman}
\caption{Two-body} \label{fig:two-body feynman}
\panel{.19}
\includepgf*{three-body-feynman}
\caption{Three-body} \label{fig:three-body feynman}
\panel{.22}
\includepgf*{four-body-feynman}
\caption{Four-body} \label{fig:four-body feynman}
\panel{.22}
\includepgf*{five-body-feynman}
\caption{Five-body} \label{fig:five-body feynman}
\panel{.21}
\includepgf*{five-body-feynman-3}
\caption{Five-body} \label{fig:five-body feynman-3}
\end{panels}
\caption[\sentence\PSlong diagrams as dual Feynman diagrams]{
The \PS diagrams can be interpreted as being dual to tree-level Feynman diagrams.
The parameterisation in terms of differential two-body \PSs is particularly well suited to describe Feynman diagram consisting of three-particle interactions.
} \label{fig:feynman}
\end{figure}

We have derived expressions for \PSs of an arbitrary number of particles.
While the differential \PS is especially simple to express in terms of invariant masses the integration in this parameterisation is unnecessary complicated, since the volume of a four-dimensional parallelotope appears explicitly and a large number of non-trivial integration limits have to be taken into account.
On the other hand a parameterisation in terms of a minimal set of invariant masses and fitting polar opening and azimuthal decay plane angles leads to the most readily integrable differential \PS.
In order to simplify the discussion of such \PSs we have developed a graphical depiction that allows to argue about high dimensional \PSs.
This graphical description differs from prior attempts \cite{Asribekov:1962tgp,Jing:2020tth}.
In particular when expressed in terms of two-body \PSs these diagrams can be interpreted as dual diagrams to tree-level Feynman diagrams consisting of three-particle vertices, see \cref{fig:feynman}.

\printbibliography

\end{document}